\documentclass[letterpaper,twocolumn,10pt]{article}
\usepackage{usenix2019_v3}

\input{packages}
\newcommand{\name}{NanoSort\xspace}

\newcommand{\steve}[1]{}
\newcommand{\theo}[1]{}
\newcommand{\greg}[1]{}

\graphicspath{{figures/}}

\definecolor{LightGray}{gray}{0.9}
\definecolor{Gray}{gray}{0.8}
\definecolor{DarkGray}{gray}{0.5}

\usepackage{listings} % needed for the inclusion of source code

\usepackage{color}

\definecolor{dkgreen}{rgb}{0,0.6,0}
\definecolor{gray}{rgb}{0.5,0.5,0.5}
\definecolor{mauve}{rgb}{0.58,0,0.82}

\usepackage[normalem]{ulem} % for strikeout
\usepackage{outlines}

\usepackage{filecontents}

\begin{document}
%-------------------------------------------------------------------------------

%don't want date printed
\date{}

% make title bold and 14 pt font (Latex default is non-bold, 16 pt)
\title{\Large \bf From Sand to Flour: The Next Leap in Granular Computing with NanoSort\\
%(Submission \#369)
}
 
%for single author (just remove % characters)
%\author{
%{\rm Theo Jepsen}\\
%Stanford University, Intel
%\and
%{\rm Stephen Ibanez}\\
%Stanford University, Intel
% \and
% {\rm Gregory Valiant}\\
%Stanford University
% \and
% {\rm Nick McKeown}\\
%Stanford University, Intel
% copy the following lines to add more authors
% \and
% {\rm Name}\\
%Name Institution
%} % end author

\author{{\rm Theo Jepsen$^{\star\dag}$, Stephen Ibanez$^{\star\dag}$, Gregory Valiant$^{\star}$, and Nick McKeown$^{\star\dag}$}\\ 
$^{\star}$Stanford University~~~$^{\dag}$Intel
}

%\author{}

\maketitle

%-------------------------------------------------------------------------------
\begin{abstract}
%-------------------------------------------------------------------------------

% Conjecture: by reducing communication overhead, it is worth farming out small pieces of compute. Smaller pieces of compute lets us reshape the execution graph of distributed applications: go ``wider'' and ``shallower'', increasing parallelism while reducing runtime. Short-lived \emph{nanoTasks}, which complete within hundreds of nanoseconds, can be tightly coupled with low-latency communication to provide \emph{nanoServices}.

%Distributed processing is limited by communication overhead (the ratio of communication time to processing time).
%\greg{``communication overhead'' isn't actually the ratio of communication time to processing time, is it???}
%If communication overhead is large, there is no point in farming out lots of very small tasks --- the communication overhead will dominate.
%We are therefore interested in seeing how fast you can make distributed processing by driving down communication overhead as far as possible.
%In doing so, we can scale out distributed processing to tens of thousands of small, nanosecond-scale, tasks that work together to provide a \emph{nanoService}.

%Driving down the mean communication time is not enough; you need to drive down the tail of the communication time; with a very distributed computation, this matters even more. We are therefore very interested in the effect of tail communication latency on the computation time, and the benefits of the low-variance properties of a new NIC/CPU hardware design, the nanoPU.

The granularity of distributed computing is limited by communication time: there is no point in farming out smaller and smaller tasks if the communication overhead dominates the decrease in processing time due to the added parallelism.
In this work, we leverage the low communication latency of a new NIC/CPU hardware design, the nanoPU, to explore a new extreme of granularity in distributed computation, where a problem is partitioned into tens of thousands of nanosecond-scale tasks.

% To demonstrate the feasibility of We want to make it
% real and practical, yet possible to explain and understand in a single
% paper. We therefore picked Sorting as our application, allowing us to
% benchmark against known results. We demonstrate sorting 1M keys over 10x
% faster than previously reported, by employing a massive number of small
% nanoPU cores.
To evaluate the feasibility and practicality of extremely fine-grained computing, we built NanoSort, a distributed sorting algorithm running on the nanoPU.
Using cycle-accurate FireSim simulations of 65,536 nanoPU cores, we show that NanoSort can sort 1M keys in 68$\mu$s, an order of magnitude faster than MilliSort, the current state-of-the-art.

\end{abstract}
%-------------------------------------------------------------------------------
\section{Introduction}
\label{sec:intro}
%-------------------------------------------------------------------------------

%With ever growing data sets and real-time latency requirements, we need to make workloads run faster. 
There are two main ways to speed-up compute workloads: use beefier CPUs with higher clock speeds; or split the workload into smaller, finer-grained pieces that run in parallel.

With the slowing of Dennard scaling, clock speed is not increasing as fast as it once did, the general trend is towards increased parallelism. Over the past twenty years, we have seen parallelism increase with multicore CPUs,from dual-core, to quad core, to 64-core CPUs. It is not uncommon to see a single server with hundreds of cores (e.g., 224 cores per server~\cite{platinum9200}).
To further increase parallelism, workloads frequently expand beyond a single server, to racks of servers connected by high-speed networks. 
As we add more cores, however, the cost of inter-core communication increases: from O(ns) within the same CPU, to O(10ns) between CPUs on the same server, and to O(10$\mu$s) across the network.

As Gene Amdahl said in 1967 ``the overall performance improvement gained by optimizing a single part of a system is limited by the fraction of time that the improved part is actually used''~\cite{amdahl1967validity}. In our context, as we break the overall workload into smaller and smaller compute tasks to be processed in parallel, we eventually run into the overhead cost of communication between the cores. If we make the tasks too small, the completion time is dominated by the communication time between CPU cores. Ultimately, if we break the job into tiny tasks, farmed out to a huge number of CPU cores, the completion time will be dominated by the communication time. The communication time limits the degree of parallelism.

We have seen this trend with one of the most executed~\cite{vitter1985bucket,klerlein1988transition} and
studied~\cite{astrachan2003bubble,aggarwal1988io,vitter1985bucket} workloads in computer science\footnote{Sorting accounted for a quarter of computing time in 1984~\cite{vitter1985bucket}.}: sorting.
Initially, the biggest speedups were achieved by scaling \emph{up} to beefier CPUs.
Then, systems began scaling \emph{out} to multiple cores.
In 1986, Tsukerman et al.\ FastSort~\cite{fastsort} sorted 1M records in 3600 seconds with 2 cores.
In 1989, it took Baugsto et al.\ 180 seconds with 16 cores, and 40 seconds with 100 cores~\cite{baugsto1990}.
In 1993, Nyberg et al.\ sorted in 7 seconds using AlphaSort~\cite{nyberg1995alphasort} with 3 cores.

When the single-server limit was reached, they started using multiple servers connected by the network.
In 2002, Popovici et al.~\cite{popovici2002datamation} sorted in 440 milliseconds with 32 servers, each with two P3 CPUs.
Then, they started using racks of servers, but optimized for throughput, not latency.
In 2009, Hadoop used 3,452 nodes with 2 quadcore CPUs each~\cite{sortbenchmark}. 
In 2016, TencentSort~\cite{tencentsort}, the current record holder of the GraySort "Daytona" (sorting throughput category) used 512 nodes with 20 cores each.
These large scale systems were trading-off \emph{latency} in favor of \emph{throughput}.
Recently, MilliSort~\cite{millisort} tried pushing the limits of large-scale, fast (low-latency) sorting, by using RAMCloud, a high performance networking subsystem.
They achieved an impressive speed-up with finer granularity: they sorted 1M records in 1ms using 2,240 cores.
But they were fundamentally limited by Amdahl's law. When MilliSort's cost of network communication (using RAMCloud with a 5$\mu$s RTT) dominated the runtime, they could no longer decrease the size of compute tasks, limiting the scale-out.

The networking community has been working to reduce the communication overhead.
Recently we have seen several designs that reduce latency, especially at the end-host, which is responsible for most of the of the latency~\cite{qizhe2021understanding}.
By optimizing the software stack using currently available network cards (NICs), eRPC~\cite{eRPC} reduces the wire-to-wire latency to 850ns.
NeBuLa~\cite{nebula} presents a hardware design that delivers packets straight from the NIC to the CPU L1 cache, with a latency of 100ns.
These systems bring two main benefits: lower overhead and reduced latency.
To reduce the mean and \emph{tail} latency, the nanoPU~\cite{nanopu} offloads the entire RPC stack to NIC hardware and provides a dedicated \emph{fast path} directly between the network and CPU register file. The nanoPU's wire-to-wire round-trip latency through a core is just 69ns with almost no additional tail latency.

%\begin{table*}[ht!]
%\centering
%\begin{tabular}{lr}
%\toprule
%\textbf{App} & \textbf{Latency Requirement} \\
%\midrule
%Render a single frame @60FPS & 16ms \\
%\multicolumn{1}{r}{(after 10ms RTT over 5G~\cite{rtt5g})} & 6ms \\
%Autonomous vehicle avoid obstacle 5m away~\cite{yu20} & 164ms \\
%% https://www.cnet.com/tech/mobile/5g-latency-why-speeding-up-networks-matters-faq/
%\multicolumn{1}{r}{(after 70ms RTT over 4G)} & 94ms \\
%Fast re-route (FRR) path calculation & microseconds \\
%\bottomrule
%\end{tabular}
%\caption{Requirements of real-time applications}
%\label{tab:latency-reqs}
%\end{table*}

Generally, we decide to stop increasing parallelism when the limits of the underlying communication system are reached.
For example, MilliSort reached this limit with 2,240 cores.
Since the nanoPU has reduced the communication latency by an order of magnitude, from microseconds to nanoseconds, it may be able to push the limit further.
We now have the opportunity to take granularity to a new extreme. This raises the question: \emph{{\textbf if we had an infinite number of cores, how much of a speed-up would we get?}}

When communication latency is on the order of microseconds, the workload takes O(ms). For example, with a RTT of 5$\mu$s and 10 communication steps, the cost will be 50$\mu$s, not even including the compute time. If the communication latency is reduced to nanoseconds, will workloads complete within microseconds?

Using a network stack with low communication overhead is essential, but not the sole requirement to enable fine-granularity and massive parallelism.
% we cannot simply pick any application and run it on the nanoPU.
An application must be amenable to parellilization, particularly, it must be possible to split the application into fine-grained tasks.

Increasing granularity to the extreme presents three main challenges. First, we have to shrink the task size as much as possible, but not too small, as it will have to pay the price of constant overheads. Also, as tasks get smaller, we need more parallel tasks, which increases the amount of communication.
Second, the tasks work dependency graphs must be structured in such a way to maximize system utilization. If tasks have too many dependencies, they may spend too much time waiting, reducing the overall parallelism.
Third, the load must be evenly distributed across cores. This is especially important for sparse workloads, which are unpredictable, so must be partitioned in such a way to reduce variance. Otherwise, imbalanced partitioning will lead to some tasks becoming stragglers.

%\steve{I think it's useful to highlight the importance of figuring out how to evenly distribute load across cores. This was a key challenge for both sorting and massively parallel ray tracing.}
%\theo{The load balancing you described is covered by the third challenge above. I rephrased it to explicitly mention load balancing.}

We believe that there are two key ideas for addressing these challenges of granular computing: reduce overhead and use efficient algorithms.
New low-latency, predictable networking stacks, are essential for reducing the overheads. This enables tasks to send and receive many small messages, enabling finer granularity.
It is also necessary to design algorithms that overcome the imbalance of partitioning the data, as well as structuring the communication, to reduce the length and duration of the critical path in the work dependency graph.

To demonstrate how these key ideas can push the limits on fine-grained computing, we explore the problem of distributed sorting.
Sorting is a good problem to study because: it is a familiar problem that is well understood and easy to describe~\cite{astrachan2003bubble}; it is industrially applicable, from analytics to graphics; and its unpredictable nature makes it non-trivial to implement as a distributed algorithm that stresses system resources~\cite{tritonsort}.

%\steve{Being ``particularly difficult'' isn't a great reason why it's a good application to study. It would be nice if we can say that we think it is representative of a larger class of applications.}
%\theo{I rephrased "particularly difficult". We believe sorting is representative, but we don't have a citation for that yet... I'm still searching}

We designed a new sorting algorithm, NanoSort. NanoSort runs on a cluster with tens of thousands of nanoPU cores using efficient communication patterns, which minimize runtime variance.
The nanoPU provides very low overhead communication with just 69ns wire-to-wire latency. Such low overhead makes it possible for NanoSort to use very small tasks, on the order of 100s of nanoseconds.
The NanoSort algorithm is recursive, so it breaks the problem into smaller sub-problems in each iteration. By fanning out quickly, the critical path (the depth in the work dependency graph) is minimized, reducing the overall runtime.
The nanoPU's predictable latency, together with NanoSort's partitioning, minimizes the variance. The predictable latency ensures tasks complete in time, while the balanced partitioning balances the work across all the tasks.

To see whether NanoSort can scale, we used hundreds of Amazon EC2 instances to run cycle level simulations of thousands of nanoPU cores. We ran the GraySort 1M sorting benchmark, which measures the end-to-end time to sort 1M records. Using 65,536 cores, NanoSort sorts 1M records in 68$\mu$s, which is an order of magnitude faster than MilliSort's 1ms record.

Overall, we make the following contributions:
\begin{itemize}[itemsep=0em, leftmargin=*]
    \item Explore the design space for distributed algorithms to efficiently use new low latency communication stacks;
    \item Introduce NanoSort, a sorting algorithm designed for low-latency, fine-grained computing; and
    \item Perform a large scale, cycle level simulation of NanoSort sorting 1M keys in 68$\mu$s, an order of magnitude faster than the state of the art.
\end{itemize}

The rest of the paper is structured as follows.
The next section provides background on existing network stacks and new low latency network stacks.
Section~\ref{sec:nanoservices} explains how workloads can be restructured for even finer-grained tasks.
Sections~\ref{sec:design} and~\ref{sec:implementation} describe the design and implementation of our new sorting algorithm, NanoSort.
Section~\ref{sec:eval} evaluates the parameters that affect fine-grained computing, as well as a large scale simulation of NanoSort.

\section{Low Latency Communication Systems}
\label{sec:background}
%-------------------------------------------------------------------------------

% Based on: https://www.usenix.org/system/files/osdi21_slides_ibanez.pdf
%\begin{table}
%\centering
%\begin{tabular}{lcc}    
%\toprule
%\multirow{2}{*}{\bf System} & \multicolumn{2}{c}{\bf Latency (ns)}\\
%% \cline{2-3}
% & \textbf{Median} & \textbf{99th~\%ile} \\
%\midrule
%eRPC & 850 & XXX \\
%NeBuLa & 100 & XXX \\
%nanoPU & 69 & XXX \\
%\bottomrule                                                              
%\end{tabular}
%\caption{Median and 99th \%ile wire-to-wire loopback latency for three systems.}
%\label{tab:latencies}
%\end{table}
\begin{table}
\centering
\begin{tabular}{lr}    
\toprule
\textbf{System} & \textbf{Median latency (ns)}\\
\midrule
eRPC & 850  \\
NeBuLa & 100 \\
nanoPU & 69 \\
\bottomrule                                                              
\end{tabular}
\caption{Median wire-to-wire loopback latency for three end-host network stacks.}
\label{tab:latencies}
\end{table}

\theo{Just like in the introduction, we should add "numbers" here. E.g., what are the overheads with vanilla linux vs. Shinjuku/Shenango}

In recent years, there has been a significant amount of research effort in both industry and academia to design low latency, high throughput communication systems.
Systems like Shinjuku~\cite{shinjuku} and Shenango~\cite{shenango} recognize the high overheads associated with the Linux network stack and opt to use a custom data-plane operating system that is designed to efficiently schedule network requests and application threads across cores.
The authors of eRPC~\cite{eRPC} demonstrate that it is possible to achieve very low latency RPCs using commodity NICs by leveraging a number of clever common case optimizations.
Many commercial NICs include support for remote direct memory access (RDMA) which enables low latency by performing zero-copy communication between application buffers on remote systems and by offloading the transport layer into NIC hardware.
Another approach to optimize communication performance is to design new datacenter transport protocols.
NDP~\cite{ndp} and Homa~\cite{homa} are two recent datacenter transport protocols that achieve low latency communication by leveraging receiver-driven solicitation, and in-network support for packet trimming and priority queueing.
One of the most promising and ambitious approaches we have seen to optimize communication performance is to integrate the CPU and NIC, thus bypassing PCIe.
NeBuLa~\cite{nebula} and the nanoPU~\cite{nanopu} are two such systems that take this approach.
The nanoPU is able to achieve a wire-to-wire latency of just 69ns, with very little variance, and each core is able to process over 100M requests per second.
\theo{why is low variance important? I think this section should emphasize the importance of tail latency, i.e. the "tail at scale" discussion}
While the nanoPU is a research prototype, we choose to use it as a platform for fine-grained computing because we believe its performance characteristics are representative of future datacenter network infrastructure.
\theo{why is it representative?}
Additionally, an open source nanoPU prototype is readily available~\cite{nanopu-github} and can be used to implement large scale application on Amazon cloud.
The following section describes some of the key characteristics of the nanoPU that allow it achieve such high performance.

\subsection{nanoPU Background}
The nanoPU is a hero experiment intended to push the performance limits of the host network stack.
It is designed to minimize average and tail latency as well as the CPU overhead associated with network communication.
By doing so, it enables applications threads on remote cores to communicate quickly and predictably at high message rates.
The nanoPU ``fast path'' uses a novel approach to deliver data directly between the network and the CPU register file - bypassing PCIe, caches, and the memory hierarchy entirely.
In order to minimize tail latency, the nanoPU fast path implements the network stack's key resource scheduling decisions in efficient hardware: (1) load balancing of network messages across cores and (2) scheduling of application threads on cores.
Furthermore, the nanoPU implements latency-optimized network transport protocols in programmable hardware thus reducing the CPU overhead associated with network communication
Additionally, the nanoPU's simple two-register interface enables applications to send and receive small messages at very high rates.

\begin{figure}[!t]
\centering
\begin{itemize}
  \setlength{\itemsep}{1pt}
  \setlength{\parskip}{0pt}
  \setlength{\parsep}{0pt}
    \item Scan 1K 8-byte words in L1 cache
    \item Sort 40 8-byte keys (Section~\ref{sec:eval})
    % \item load XXX bytes from L2 cache
    % \item load XXX bytes from memory
    \item Travel 300m at the speed of light 
    \item Receive 2Kbyte on a 200Gbps NIC
    \item Process 118 8-byte loopback nanoRequests~\cite{nanopu}
    \item Perform 2 MICA R/W operations~\cite{nanopu}
    \item Perform 4 set algebra intersections~\cite{nanopu}
\end{itemize}
 \vspace{-0.5pt}
    \caption{Examples of what can be accomplished in under 1 $\mu$s (computations using a nanoPU core).}
    \label{fig:under1us}
\end{figure}

\section{Extremely Granular Computing}
\label{sec:nanoservices}

Current systems have tasks (e.g. RPCs) that run for hundreds of microseconds~\cite{microservices,killer-microseconds,deathstarbench,shenango,shinjuku}. One reason for long-lived tasks is to justify the communication cost of invocation: if it takes microseconds~\cite{eRPC} to invoke the task, it should run for microseconds to amortize the communication cost.

As the communication latency and overhead decrease, it is worth farming out smaller, shorter-lived tasks. With the 69ns wire-to-wire communication latency of the nanoPU, tasks only need to run for hundreds of nanoseconds to amortize the communication..

A time budget on the order of hundreds of nanoseconds may seem too short to perform useful work for an application, especially on a general purpose CPU.
Figure~\ref{fig:under1us} shows examples of common operations and tasks that complete within 1$\mu$s on a 3.2GHz in-order RISC-V Rocket core.
When data is cache resident, which is the case for many of those operations, most of the time is spent on ``useful'' compute cycles, instead of waiting for memory I/O.
Alone, a single small task may not be able implement a meaningful application.
However, together, thousands of fine grained tasks can execute a large job, harnessing the aggregate CPU cycles and cache bandwidth.

There are two main benefits of using fine-grained tasks. First, by breaking a large job into smaller pieces, it is possible to increase parallelism. It lets us re-shape the compute core-hours: we can run the same number of core-hours in less time, going from ``deep'' to ``wide''. Second, the compute resources can be scaled to the precisely match the size of the job: we can allocate compute at the granularity of cores or threads, instead of servers or entire racks.

Parallelism doesn't come ``for free'' though. Job completion time depends on the depth of the task dependency graph, which is a graph of the communication pattern between the tasks defined by the algorithm (e.g., Figure~\ref{fig:mergetree}).
The deeper the graph, the longer the execution. To reduce the execution time, the graph can be reshaped to be shallower and wider. Wider means more parallelism is possible, which requires smaller compute tasks.
In turn, it requires more communication, so it is essential that the communication overhead be low. The width and depth of the dependency graph determine the amount of communication as well as the size of the tasks.

\subsection{A Simple Example: MergeMin}
\label{sec:minmerge}

\begin{figure}[t]
    \centering
    % \captionsetup{justification=centering}
    \begin{subfigure}[t]{.6\columnwidth}
        \centering
        \includegraphics[width=\columnwidth]{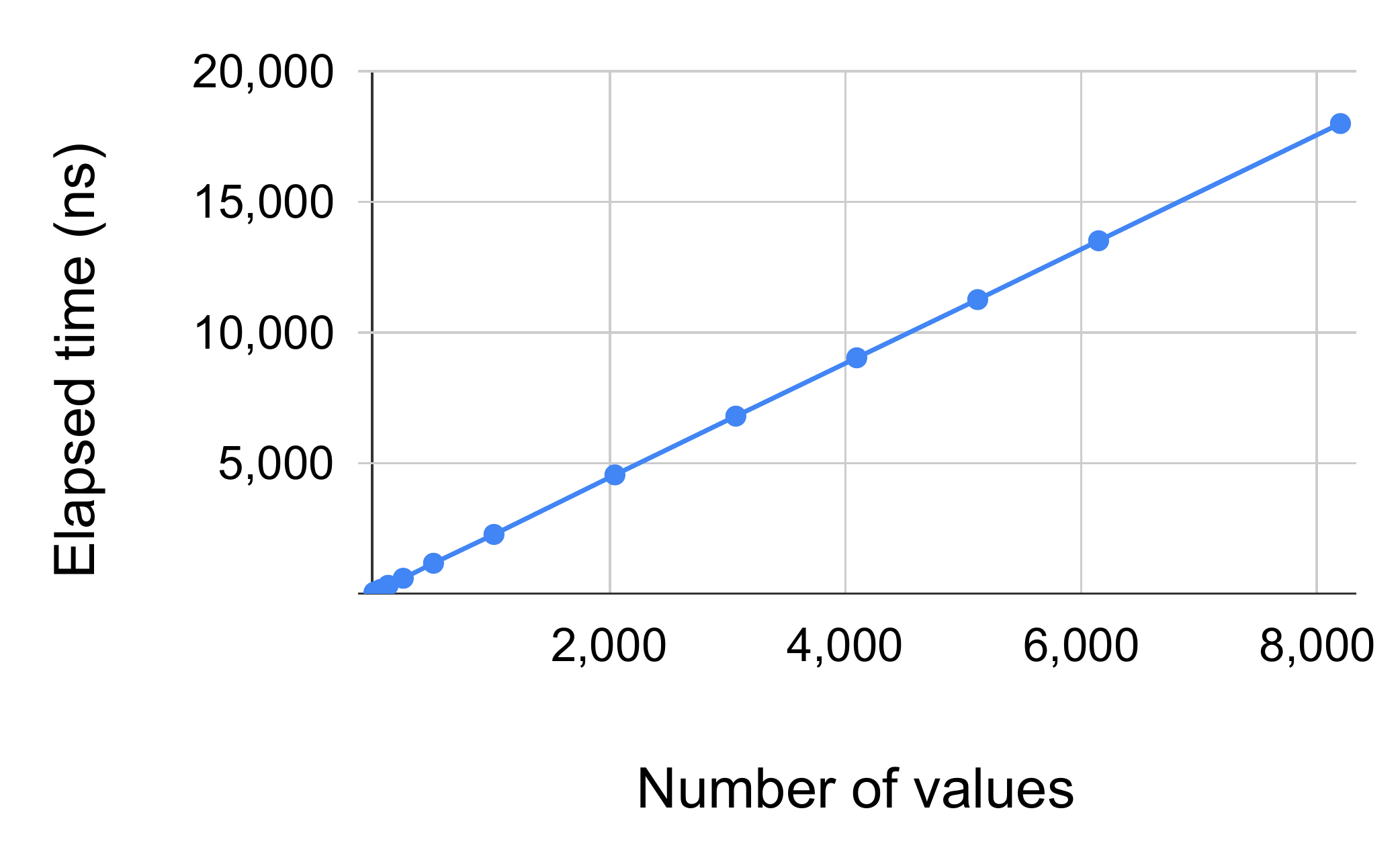}
        \vspace{-15pt}
        \caption{Runtime}
        \label{fig:minmerge-time}
    \end{subfigure}
    \begin{subfigure}[t]{.6\columnwidth}
        \centering
        \includegraphics[width=\columnwidth]{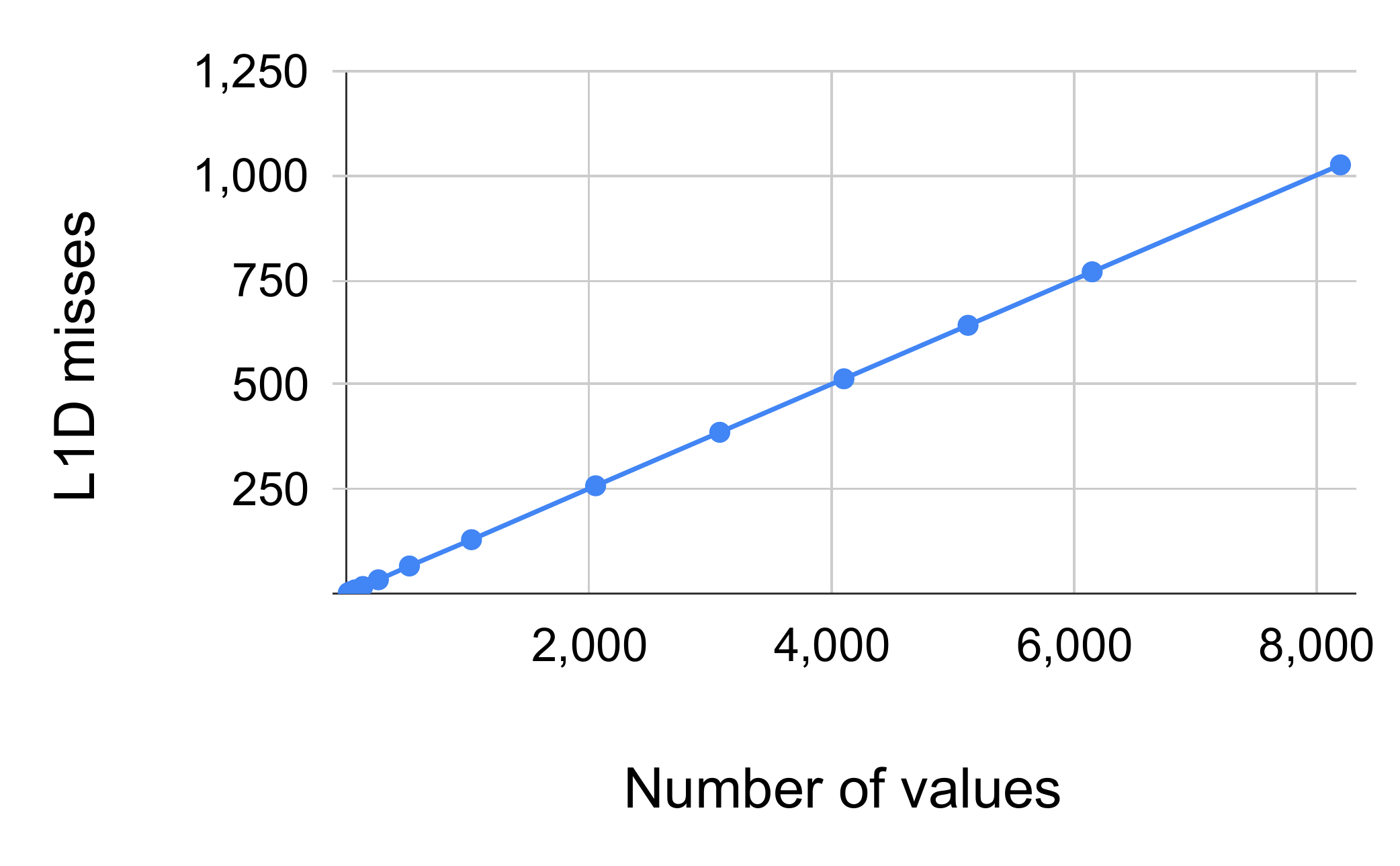}
        \vspace{-15pt}
        \caption{Cache misses}
        \label{fig:minmerge-misses}
    \end{subfigure}
    \caption{Using a single core to find the minimum from an increasing number of values.}
    \label{fig:minmerge-local}
\end{figure}

To illustrate the performance implications of a dependency graph's width vs. depth, we present a simple example: finding the minimum value from a list of numbers. This could be done sequentially, scanning the numbers while keeping a running minimum.
However, as we can see in Figure~\ref{fig:minmerge-local} this is limited by the processing speed of a single core, which in turn is limited by the I/O throughput, or cache misses (Figure~\ref{fig:minmerge-misses}).

\begin{figure}[t]
    \centering
    \includegraphics[width=1\linewidth]{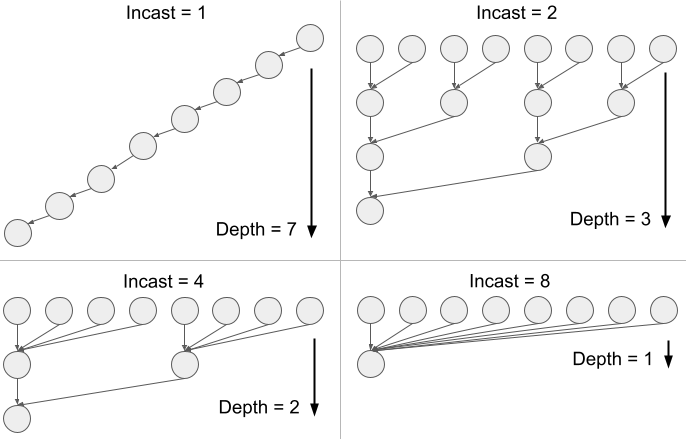}
    %\vspace{-20pt}
    \caption{Lower incast results in deeper work trees}
    \label{fig:mergetree}
\end{figure}

By using more workers, this process can be parallelized. It cannot be fully parallelized, however; it requires some communication between the workers to merge their minimum values. This can be done using a merge tree, as shown in Figure~\ref{fig:mergetree}.
At each level in the tree, a worker receives the minimum value of one or more other workers; in turn, the worker merges these minimums, and passes it down the tree.

\begin{figure}[t]
    \centering
    \includegraphics[width=0.8\linewidth]{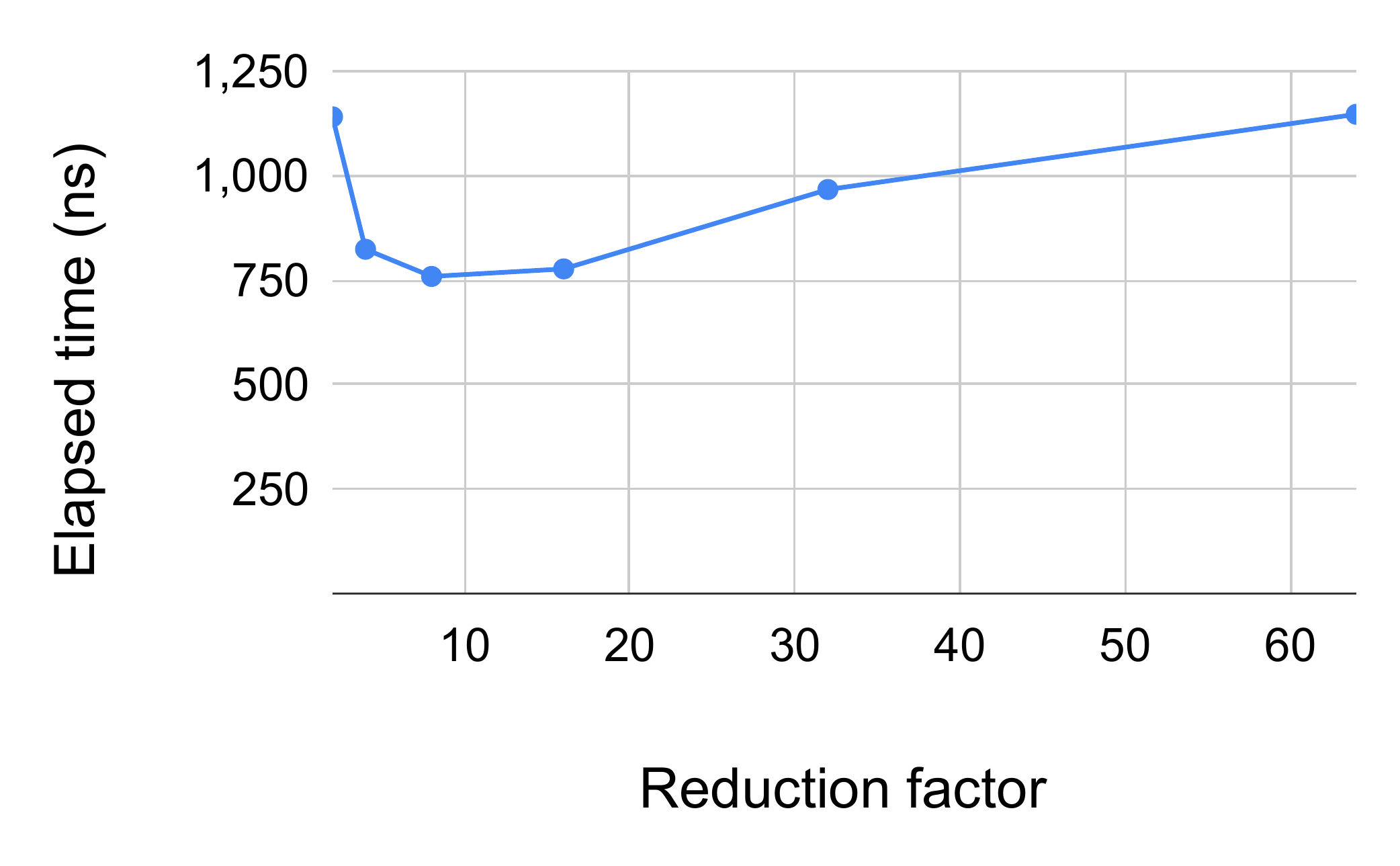}
    %\vspace{-20pt}
    \caption{The runtime for MergeMin with increasing incast size (64 cores, 128 values per core).}
    \label{fig:minmerge-incast}
\end{figure}

This is repeated until reaching the root of the tree, where a single worker merges the final minimum values. There are two extremes of this merge tree: either very deep, or very shallow. This determines the ``incast'': the number of values that are received and merged at each level in the tree.
A very shallow tree has large incast. This increases the latency for each level of tree, because the merger worker has to receive and process more minimum values. A deep tree has smaller incasts, so each level is faster, but there are more levels.
As we can see in Figure~\ref{fig:minmerge-incast}, there is a trade-off between width and depth, which has sweet spot with an incast size of 8.

The runtime depends on the choice of two parameters: the granularity (i.e. the amount of work per worker, the inverse of the number of workers), and the depth of the tree (determining the incast size). The two parameters cannot be chosen independently. For example, if the granularity (number of workers) is increased, the tree becomes wider. This will cause the tree to become deeper, unless the incast size is increased.

\steve{Perhaps another way of saying this is that the two parameters that determine the optimal balance between width and depth are: (1) the rate at which a single core can process network messages, and (2) the tail communication latency between cores. We are seeing significant improvements in both of these with recent network systems (e.g. nanoPU) and thus we need to revisit how this affects optimal algorithm design.}

\subsection{Building Applications with Fine-Grained Tasks}

Although many applications are amenable to being implemented with granular computing, applications that have real-time latency requirements benefit the most from fine-grained tasks.
In some cases, such as interactive systems, it is more important to provide a result in a timely fashion, rather than optimize resource utilization (efficiency, throughput).
Applications that have these requirements, and that are highly parallelizable, are the best suited for granular computing.

Fine-grained computing is an especially good fit for interactive applications, like real time data analytics and graphics processing. Stream processing systems like Spark ingest large amounts of data and must quickly produce a result. Many of its basic operations, like mapping (transforming) and reducing (merging) data are easily parallelizable. Graphics processing, like ray tracing, also has many operations that can be parellelized. For example, constructing the bounding volume hierarchy (BVH) requires sorting the objects in the scene. Web search also has interactive requirements, and can be executed with fine-grained tasks using set algebra operations~\cite{nanopu}.

Many of these applications exhibit the Map Reduce~\cite{mapreduce} pattern, which is is a natural fit for granular computing. The first step is to transform the input data, which can generally be done independently, in parallel, by all the nanoTasks running on different cores. The reduce step, however, requires coordination among the cores. The design of the reduce (merge) has important implications for how the granular program is implemented.

It is essential not only to minimize the size of individual tasks, but also the critical path through the dependency graph. This requires the granular program to follow some design principles:
\begin{itemize}[itemsep=0em, leftmargin=*]
    \item Asynchronous communication. Tasks should not block waiting for responses after sending out a message. It is better to use the ``fire and forget'' communication model and build synchronization into the algorithm.
    \item Decentralized decision-making. Workers should make decisions without a global coordinator, which could become a bottleneck in the task dependency graph. This also allows multiple threads to progress through the dependency graph in parallel.
    \item Tiny tasks. Each task should be as small as possible and entirely cache-resident.
    \item Even load distribution. In expectation, all tasks should process the same amount of data and send/receive the same amount of messages, independently of the workload.
\end{itemize}

Some of these guiding principles may seem obvious, but there are some nuances when actually implementing them. In the next section we discuss how they can be applied to a specific application: sorting.

\section{\name Design}
\label{sec:design}

Over the years, the requirements and constraints faced by sorting algorithms have continued to evolve~\cite{vitter1985bucket,klerlein1988transition,astrachan2003bubble,aggarwal1988io}.
In the early days, sorting was CPU bound, which was suitable for algorithms that do arbitrary comparisons (e.g., InsertionSort, BubbleSort, QuickSort).
Then, it was limited by disk, so external sorting algorithms were designed to optimize I/O (e.g., reducing disk seeks~\cite{aggarwal1988io}).
More recently, as sorting has scaled out to clusters~\cite{tritonsort,tencentsort}, distributed systems have adopted various forms of BucketSort~\cite{vitter1985bucket}.

\theo{Should we refer to this class of algorithms as BucketSort or distribution sort?}

MilliSort is a bucket based sorting algorithm that minimized network communication by dividing the algorithm into two parts: partitioning and shuffling.
During partitioning, a subset of the nodes pick bucket boundaries, which are disseminated to all other nodes.
During the shuffle, the boundaries are used to route records to the final node which is responsible for a bucket.

These distributed sorting algorithms like MilliSort were designed to minimize communication, in order to avoid network overhead.
Now that systems like the nanoPU provide low-overhead communication, it is no longer necessary to \emph{minimize} communication.
It may be actually beneficial to communicate \emph{more}.
In the next section we describe the high-level design of such an algorithm that takes advantage of cheap communication.

\subsection{Extremely Granular Sorting}
There is no black-box way to convert an existing algorithm into one that can be efficiently run on the nanoPU; to some extent, the algorithm must be designed from scratch with an eye towards balancing the per-node compute and communication times. \theo{Great point about finding the compute/communication balance. Did we find balance? How did we find the balance? We don't discuss this again in this section? (we use the word balance, but to refer to bucket variance, so maybe we should not use the same word for these two different concepts)} \greg{No idea if we found the balance or not---but the discussion in section and section 4.1 tell people the important considerations if they want to design a system with the right balance...I'm not sure which knobs we actually experimentally tried turning, but I don't think we should present this as a super-optimized algorithm.}
Nevertheless, this algorithm design problem need not be daunting.  In the case of nanoSort, we arrived at a simple and natural modification of quicksort, guided by the following the general design principles of minimizing unnecessary communication and exposing high-level ``dials'' (or ``knobs'') that tune the tradeoff between compute-per-node and communication.
\theo{Is NanoSort a variant of QuickSort? I think it's a type of distribution sort, specifically bucket sort: \url{https://en.wikipedia.org/wiki/Sorting_algorithm#Distribution_sorts}}\greg{I think its better to describe nanosort as a variant of quicksort--the key to both nanosort and quicksort is how the pivots are chosen....i had never heard of 'distribution sort' before seeing that wikipedia article.} 

We now describe nanoSort. For the sake of simplicity and clarity, throughout we assume that the set of keys are all distinct.  By design, all operations in nanoSort are comparison-based, and hence the execution depends only on the \emph{relative} ordering of the keys, as opposed to their actual values.  For this reason, the runtime will not depend on the distribution of the underlying keys (e.g. uniformly distributed, vs clustered/heavy-tailed, etc.).

NanoSort is a clean, recursive sorting algorithm that is a variant of quicksort.   At each level of recursion, the goal will be to split up the set of keys into $b$ ``buckets'' such that: 1) all the keys in bucket $i$ are less than those in bucket $i+1$, and 2) the number of keys assigned to each bucket is roughly equal for all buckets.   We will then recurse on each bucket.  %For each bucket, at each level of recursion, we will have a predefined set of nodes that will be responsible for those keys, and 
Crucially, after a given step of recursion, no communication will be required between nodes corresponding to different buckets.  

To motivate the specific design decisions (beyond simply setting the parameters such as the number of buckets at each level of recursion, $b$, and the average number of keys per node), we briefly discuss key considerations.

\subsection{Balanced Buckets and Balanced Nodes}  As with any variant of quicksort, one hopes that at each step of recursion, the number of keys assigned to each bucket is fairly uniform, since larger buckets would require either deeper recursion or more time per step of recursion.
%\theo{do large buckets govern \emph{depth} of recursion? I thought the number of levels of recursion was independent of bucket size variance. I'd say that variance increases runtime (tail), but not the depth.}\greg{I changed the wording, is this clear now?}
%\theo{Yes, it's clear now.}
Beyond this, in our distributed setting, if some buckets are much larger than others, we would need to ensure that the number of nodes allocated to a given bucket is proportional to the number of keys in the bucket.  If that were not the case, if each of the $b$ buckets corresponded to the same number of nodes, then nodes corresponding to buckets with more keys would end up incurring significantly more incasts.  Combined with the greater number of recursive steps, this would be especially debilitating.  Of course, to ensure that a bucket is allocated a proportional number of nodes would require counting (exactly, or approximately via sampling) the number of keys in each bucket.

To avoid these complications that would come with significant variation in the bucket sizes at each level of recursion, we opt to expend slightly more communication to ensure fairly uniform buckets.  This is achieved via a two-pronged approach.  First, we ensure that the  bucket sizes will be tightly concentrated about their expectations---namely that we will not expect much variation due to randomness in the execution of the algorithm---and then we are careful to ensure that the expected size of all buckets are equal at a given step of the recursion. The first step is accomplished via a ``median-tree'' that approximates the median while being fairly communication efficient.  The second step  uses no additional communication but requires some care. We describe these two components below.

\theo{in this two-pronged approach, what's the difference between \emph{expectation} and \emph{concentration around expectation}?}
\greg{I just meant that you want the expectation to be small, and the ``tail'' to also be close to the expectation....``concentration'' is a specific technical term--will the audience not be familiar with this usage?} 
\theo{we cannot assume the audience is familiar with the specific technical definition of ``concentration''.}
\steve{I think it's important to highlight this clever use of extra communication to achieve even load distribution. Perhaps this is a key design principal that should be mentioned in the previous section?}

For concreteness, suppose our goal is to define $b=10$ buckets by selecting $b-1$ ``pivots'', $p_1 \le p_2 \le \ldots \le p_9$.  A naive and communication-efficient way to select these pivots is simply to select 9 pivots uniformly at random (without replacement) from the set of keys.  This scheme has the fortuitous property that the expected quantile of $p_i$ is exactly $i/10$ (for example, in expectation, $p_2$ will be larger than $20\%$ of the keys).  The variance of the quantiles of these pivots (and hence the variance of the corresponding bucket sizes), however, is problematic. Indeed, there is a good chance that the largest of the resulting buckets would have more than twice as many keys as the smallest.  

This issue is easy to fix in a communication-cheap manner via a median-tree.  Suppose we have $N$ nodes that each receive 9 uniformly random keys, and let $p_i^j$ denote the $i$th smallest key received by the $j$th node.  If we define $p_i = median(p_i^1,\ldots,p_i^N),$ then the variance of the quantile of $p_i$ is $O(1/N)$, corresponding to a standard deviation of $\approx 1/\sqrt{N}.$  This could be made arbitrarily small by choosing $N$ to be large, though at the expense of significant communication in the computation of the median.  Instead of computing the median exactly, we approximate the median via a median tree: for example, if $N=10^2,$ we can partition the $N$ nodes into sets of 10, and have each set of 10 report their medians to a designated node, which then aggregates the 10 reported medians (one from each set) by taking the median.  Similarly, for $N=10^3$, a similar mechanism will result in the median-of-medians-of-medians, requiring only 3 rounds of communication.  This sort of trick maintains much of the accuracy of the overall median, while reducing the communication from linear in $N$, to logarithmic.
The issue with the above strategy, is that the expected sizes of the buckets is no longer equal.  If a node receives 9 keys drawn uniformly at random from the pool of keys, it is the case that the expected quantile of the smallest received key is $10\%$.  If, however, we have $N$ nodes that each receive 9 keys drawn uniformly from the set of keys, then the expected quantile of the \emph{median} of the smallest keys that each node received is not $10\%,$ but $\approx 7.5\%$.  Phrased differently, if we let $D$ denote the distribution corresponding to the quantile of the smallest key in a set of $9$ keys chosen uniformly at random from the entire set of keys, then the \emph{expectation} of $D$ is exactly $10\%$, but the \emph{median} of $D$ is $\approx 7.5\%$.  This discrepancy between $10\%$ and $7.5\%$ means that the bucket corresponding to keys that are at most $p_1$ would be $25\%$ smaller than we would hope.  After multiple rounds of recursion, this  discrepancy is compounded multiplicatively with each recursion.

\begin{figure}[t]
\centering
    \includegraphics[width=1.0\linewidth]{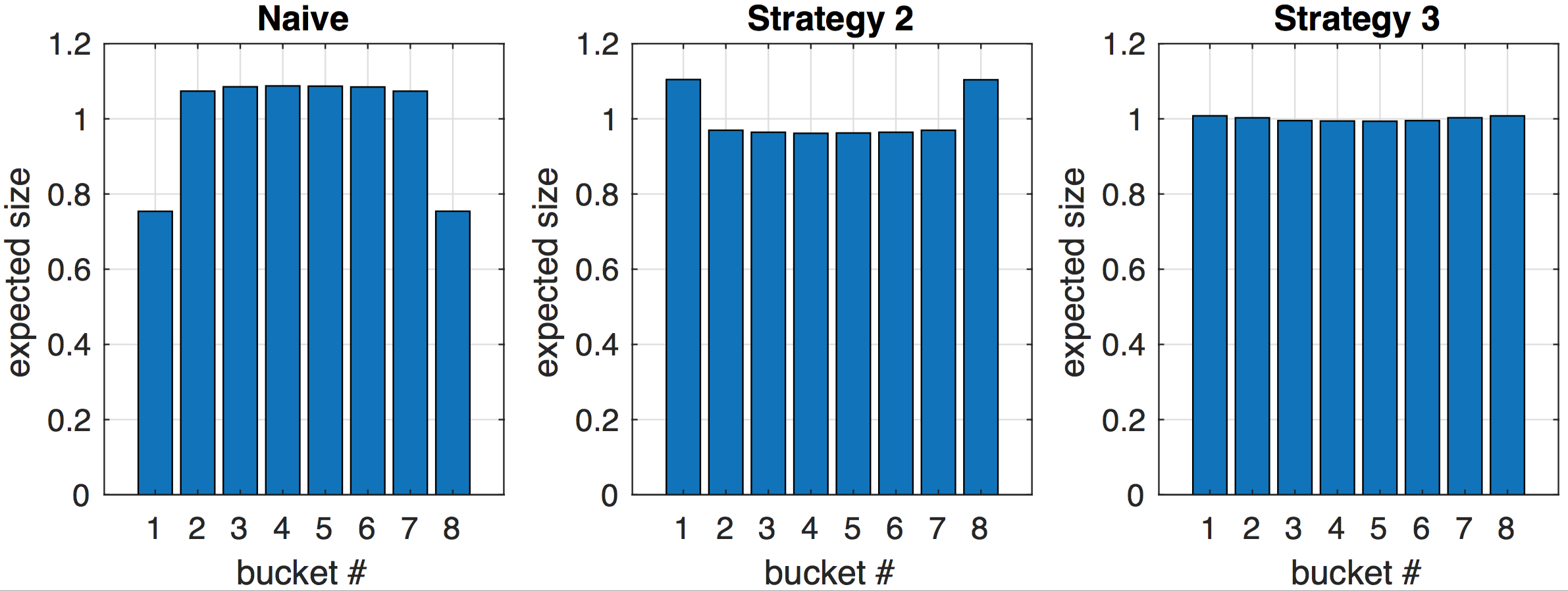}
    %\vspace{-20pt}
    \caption{Comparison of the expected bucket sizes for three pivot selection strategies in the setting where the number of buckets is 8, and the number of received keys is also 8. Naive strategy: Select 7 pivots uniformly without replacement from the 8 keys. Strategy 2: Sort the keys $k_1 \le k_2\le \ldots \le k_8$, and with probability return $k_1, k_2,\ldots,k_7,$ and with probability $1/2$ return $k_2, k_3,\ldots,k_8$. Strategy 3: Randomize between the previous two strategies as follows: With probability $1/4$ return the pivots selected by the ``Naive'' strategy, and with probability $3/4$ return the pivots selected by ``Strategy 2''.}
    \label{fig:pivotselect}
\end{figure}

To fix this issue, we simply need to take into account that we are concerned with the median, rather than the expectation.  For example, suppose $N$ nodes each receive 9 keys drawn uniformly at random from the set of keys, and the $i$th node sets $p_1^i$ to be its smallest key with probably $0.8$ and its \emph{second smallest} key with probability $0.2$, then the expected quantile of the median of these $N$ numbers is $\approx 10\%$.   Analogous modifications can similarly be made for the remaining buckets.  In this way, our algorithm is able to ``fix'' the expected quantiles of all the bucket sizes in this manner. %, while having each node randomize between two different choices for each $p_{i,j}$.  
Figure~\ref{fig:pivotselect} illustrates the expected bucket sizes for the naive pivot selection, and two other protocols. 

The PivotSelect routine described below gives our specific instantiation of this approach for the setting we implemented where the number of buckets is 16.

\begin{algorithm}
\textbf{PivotSelect (16 Buckets)}\\
Input parameters: $n=number\_keys\_at\_node,$ $num\_buckets.$  \\
Output: 15 pivots $p_1 \le p_2 \le \ldots \le p_{15}.$
\medskip
\begin{enumerate}
    \item Sort the $n$ keys $k_1 < k_2 < \ldots < k_n$.
    \item If $n=16:$ With probability $1/4$ select 15 pivots uniformly at random without replacement from $k_1,\ldots,k_{16}$ and return them (sorted).  With probability $3/8$ return $k_1,k_2,\ldots,k_{15}$, and with probability $3/8$ return $k_2,k_3,\ldots,k_{16}$.
    \item If $n<16$: Select $16-n$ keys uniformly at random from $\{k_1,\ldots,k_n\}$ and duplicate them to form a list of 16 keys, then select pivots as in the $n=16$ setting.
    \item $n \in \{17,18,\ldots,31\}:$ Select a subset of 16 keys uniformly at random, and run the $n=16$ protocol.
    \item $n=32:$ With probability $1/2$ return the keys indexed by the set $[1,3,6,8,10,12,14,16,18,20,22,24,26,27,29]$, and with probability $1/2$ return the keys indexed by the set $[4,6,7,9,11,13,15,17,19,21,23,25,27,30,32].$
    \item $n>32:$ Select a subset of 32 keys uniformly at random, and run the $n=32$ protocol.
\end{enumerate}
\end{algorithm}

Below we provide a more formal description of nanoSort.  We will require the number of nodes at our disposal to be a power of the number of buckets that is used at each level of our quicksort-like recursion.

\begin{algorithm}
\textbf{NanoSort}\\
Input parameters: $num\_keys,$ $num\_buckets,$ $num\_nodes = num\_buckets^r$. \\
\medskip

Initial Step---Random Shuffle: After this step, the keys will be randomly partitioned among the nodes subject to each node having $num\_keys/num\_nodes$ keys.

\begin{enumerate} 
\item If $num\_nodes = 1$ then sort the keys \greg{Do we need to summon ``values''?  Not sure if we are just trying to sort the keys or if we are in some setting of keys and values....Theo, please change this appropriately.}
\theo{I think summoning the values is an implementation detail. I just now described this in the implementation section. I think we should only discuss sorting keys in this section. What do you think?}
\item Otherwise (ie $num\_nodes =num\_buckets^r$ for $r \ge 1$):
\begin{itemize}
    \item For each node $i$, it sorts its keys, and extract a list of $b-1$ pivots $p^i_1\le p^i_2 \le \ldots \le p^i_{b-1}$ from its set of keys via the PivotSelect routine.
    \item Pivots $p_1,\ldots,p_{b-1}$ are selected via $b-1$ median-trees, where the $j$th median tree aggregates the $p^i_j$'s computed by each node to compute $p_j$.  The in-cast (ie fan-in) and depth of the median-tree are chosen appropriately to balance the compute time and the variance-reduction.
    \item The pivots $p_1,\ldots,p_{b-1}$ define $b$ buckets (the first bucket consists of values less than $p_1$, the next consists of values between $p_1$ and $p_2$, etc.)  The pivots are broadcast to the nodes, and the nodes are partitioned into $b$ equal sized sets (the first $num\_nodes/b$ nodes are in one set, etc).  Every node sends each of its keys to a uniformly random node in the appropriate bucket: for each key, if it is in bucket $i$, it is sent to a node chosen uniformly at random from the nodes in the $i$th partition.
    \end{itemize}
\item Recursively apply the algorithm (starting at Step 1) to each of the $b$ buckets.
\end{enumerate}
\end{algorithm}

\section{Implementation}
\label{sec:implementation}

We implemented granular computing applications, including NanoSort, as nanoPU programs. In this section we describe how we implemented the programs and the simulation cluster that we use in our evaluation (Section~\ref{sec:eval}).

\subsection{nanoPU FireSim Simulation}

The nanoPU is a SoC that extends the open-source RISC-V Rocket-Chip SoC generator~\cite{rocket-chip}.
The Rocket core is a simple five-stage, in-order, single-issue processor.
We use the default Rocket core configuration: 16KB L1 instruction and data caches, a 512KB shared L2 cache, and 16GB of external DRAM memory.

We use the Verilator~\cite{verilator} cycle-accurate simulator running on top of FireSim~\cite{firesim}.
FireSim enables us to run large-scale, cycle-accurate simulations with tens of thousands of nanoPU cores using AWS EC2 instances~\cite{ec2}.
We simulate a target clock rate of 3.2GHz---all reported results are in terms of this target clock rate.
The simulated servers are connected by switch models implemented in C++ running on the AWS x86 host CPUs.

We used the same network topology for all experiments, except for where noted otherwise. We used a two layer full-bisection topology with 200Gbps links. Each leaf switch has 64 downlinks to nanoPU NICs, and 64 uplinks to core switches. The link latency is 43ns and the switching latency is 263ns.

For our large-scale simulation, we ran 65,536 nanoPU cores using 264 EC2 instances. Of the instances, 256 simulated the nanoPU and leaf switches, while 8 instances simulated the spine switches.

The original FireSim uses AWS F1 instances to accelarate the simulation with FPGAs.
Running a simulation at this scale with AWS F1 instances would be prohibitively expensive.
We found that the NanoSort execution time is so short, that it is not worth the overhead of provisioning and flashing the FPGA.
Instead, we used cheaper general-purpose AWS compute instances running the Verilator software-based simulator.
This required significant changes to FireSim, which we are contributing back to the community.

\subsection{Applications}
We implemented microbenchmarks, MilliSort and NanoSort programs in C using the nanoPU's register-based communication API.
All the programs rely on common code for initialization and communication, as well as for the sort benchmark.

\paragraph{Communication}
A common property of both algorithms is that they have ``phases'' or ``steps''. Cores do not progress through the steps in lockstep, so cores may be in different steps from each other at the same point in time.
Thus, a core may receive messages for a subsequent step before it has completed its current step.
This requires \emph{reordering} messages, so that messages for the current step are processed before the subsequent step.

Ideally the NIC would provide a reordering mechanism to ensure that messages for the same step are delivered together.
This behaviour could be approximated by associating each step with a different L4 port, which would associate different steps with different NIC RX queues. This may lead to poor RX queue utilization. Instead, we implemented a reordering buffer in software.

\paragraph{Sort benchmark}
We perform the sorting according to the GraySort benchmark~\cite{sortbenchmark}. Just like in MilliSort's implementation~\cite{millisort}, the cluster is pre-loaded with the sorting records before starting the benchmark. The benchmark specifies that each record be 100 bytes (a 10-byte key and a 90-byte value). In order to have all the data aligned to 8-byte words for the RISCV architecture, we deviate slightly from the specification by using 104-byte records (8-byte key and 96-byte value). The data on each core is structured with all the keys in a contiguous block of memory, followed by all the values, in corresponding order.
When cores exchange keys, they also include the original core ID of each key. This way, after all the keys are sorted, the cores know where to find the original record (including the value) for each of  their keys.
At the end of the benchmark, the records are redistributed across the cores in the cluster, with the keys and values in sorted order in separate, contiguous memory regions.

\subsection{Multicast Support}
Although current commodity switches support multicast replication, they do not ensure \emph{reliable} delivery. Naively layering reliability on top of multicast is not so straight forward though, as it leads to scalability problems such as the ACK implosion problem~\cite{reliable-multicast-survey}, where the sender is flooded with an incast of acknowledgement messages. However, promising results in recent efforts~\cite{switchml} on in-network aggregation of ACKs suggest that \emph{efficient} reliable multicast is a not-so-distant reality.

Thus, since we believe it is reasonable to assume that the network can provide reliable multicast, we extended FireSim's software switch to provide reliable multicast. When a switch receives a multicast packet, it caches it before replicating it to all members of the multicast group. The cached packet can then be sent in response to a NACK or a timeout.

\section{Evaluation}
\label{sec:eval}

We evaluate the conjecture that with lower communication overhead, it is worth farming out small pieces of compute. Our main research questions are:
\begin{itemize}
    \item With low communication overhead, what are the smallest tasks possible?
    \item What is the importance of network characteristics, like latency, tail latency, and group communication?
    \item Can large-scale extreme granular computing speed-up an application like sorting?
\end{itemize}

We first run some microbenchmarks to measure the effect of parameters on the cost of performing small operations locally, as well as sending and receiving messages. Then, we show how these parameters are used in the design space of sorting algorithms.

\subsection{Microbenchmarks}
\label{sec:microbench}
In this section we explore the parameters that are necessary for minimizing the granularity of tasks. This includes the cost of local compute, as well the overhead for various communication patterns. These determine the ratio between compute and communication.
We use this to answer the question: how does the compute/communication ratio affect application runtime?

We ran experiments to evaluate the MergeMin example described in Section~\ref{sec:minmerge}. Examining this simple example helps us understand the parameters in the design space, before examining the more complex NanoSort algorithm in the next section.

First, we measure the individual parameters used for the compute ratio: we look at the runtime for local operations (finding the minimum value and sorting), and communication operations (receiving/sending messages).

%\paragraph{Local operations}

Figure~\ref{fig:minmerge-time} shows the time for a single RISC-V core to find the minimum value from a list of integers. The min operation is not compute bound (it consists mostly of comparison operations), but it is limited by the I/O, as we can see from the cache miss rate in Figure~\ref{fig:minmerge-misses}. It takes 18us to find the minimum of 8,192 values. This is clearly too long for a nanoTask that should complete within hundreds of nanoseconds.

By using many cores in parallel, we can leverage the aggregate I/O from all the cores' caches. Ideally, with 64 cores, it should take $\frac{1}{64}^{th}$ of 18us (281ns) to find the minimum value. However, this does not include communication. When structuring communication using the merge tree described in Section~\ref{sec:minmerge}, we must consider the overhead for each core to receive messages.

\begin{figure}[t]
\centering
    \includegraphics[width=0.8\linewidth]{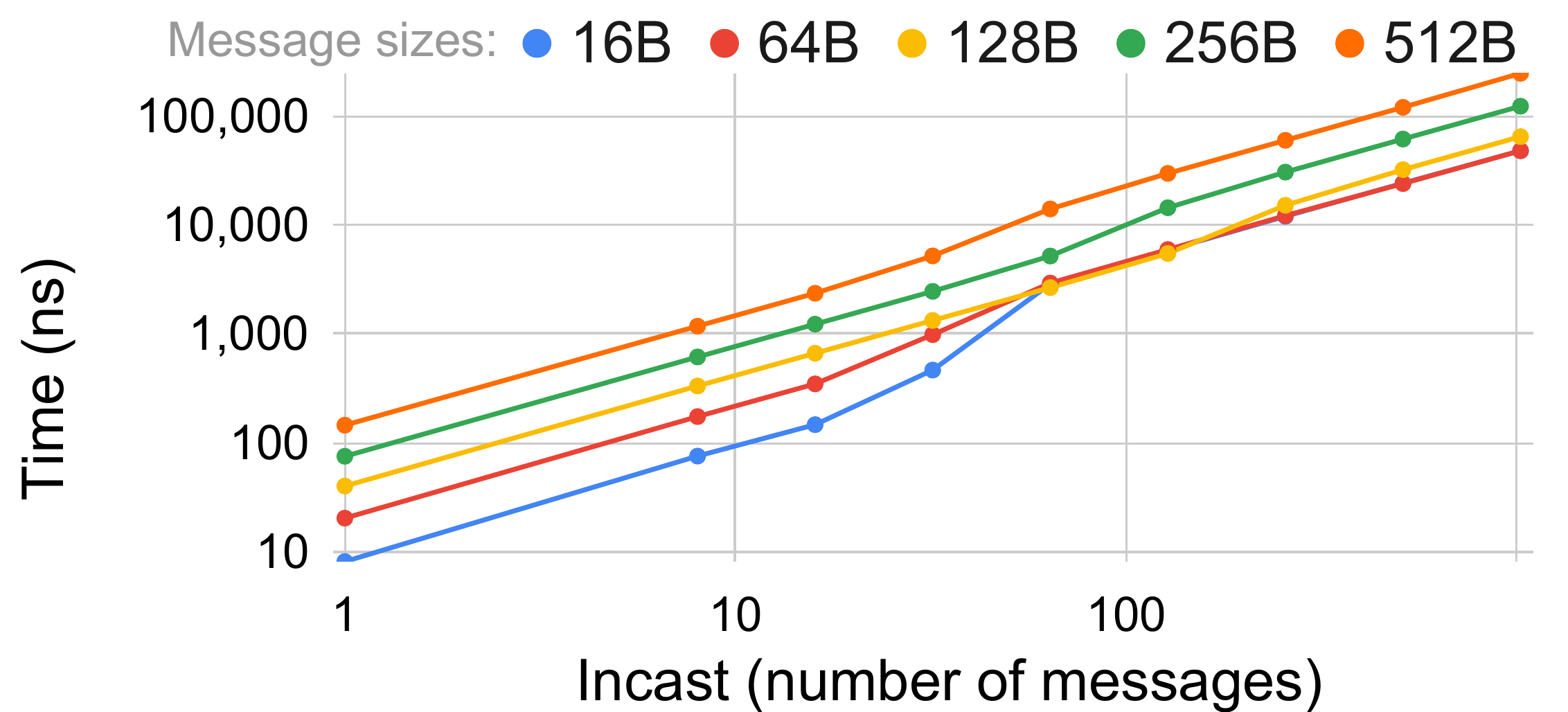}
    %\vspace{-20pt}
    \caption{Time to receive increasing number of messages.}
    \label{fig:bench-recv}
\end{figure}

Figure~\ref{fig:bench-recv} shows the time for a single nanoPU core to receive an increasing number of messages of various sizes. It takes about 8ns to receive a single 16-byte message, and 400ns to receive 64 messages. If one core were to receive and the min values from 63 other cores, it would take about 400ns.

Figure~\ref{fig:minmerge-incast} shows that there is a sweet spot for the number of values to merge at each level of the merge tree. At one extreme, with an incast of 1, the tree reduces to a straight line (see Figure~\ref{fig:mergetree}) where the runtime is dominated by propagation delay. At the other extreme, with an incast of 64, the runtime is bottle-necked by a single core receiving and merging all the values. With an incast of 8, the tree has two levels, which is the right trade-off between tree width and tree height: it finds the minimum value in 750ns.

Although MergeMin is a simple example, it provides insights into the design space and the parameters that affect performance. It is important to carefully find the appropriate balance of communication and per-core computation. We must keep this in mind when considering the design of more complex applications, as we will describe in the next section with sorting.

\begin{figure}[t]
\centering
    \includegraphics[width=0.8\linewidth]{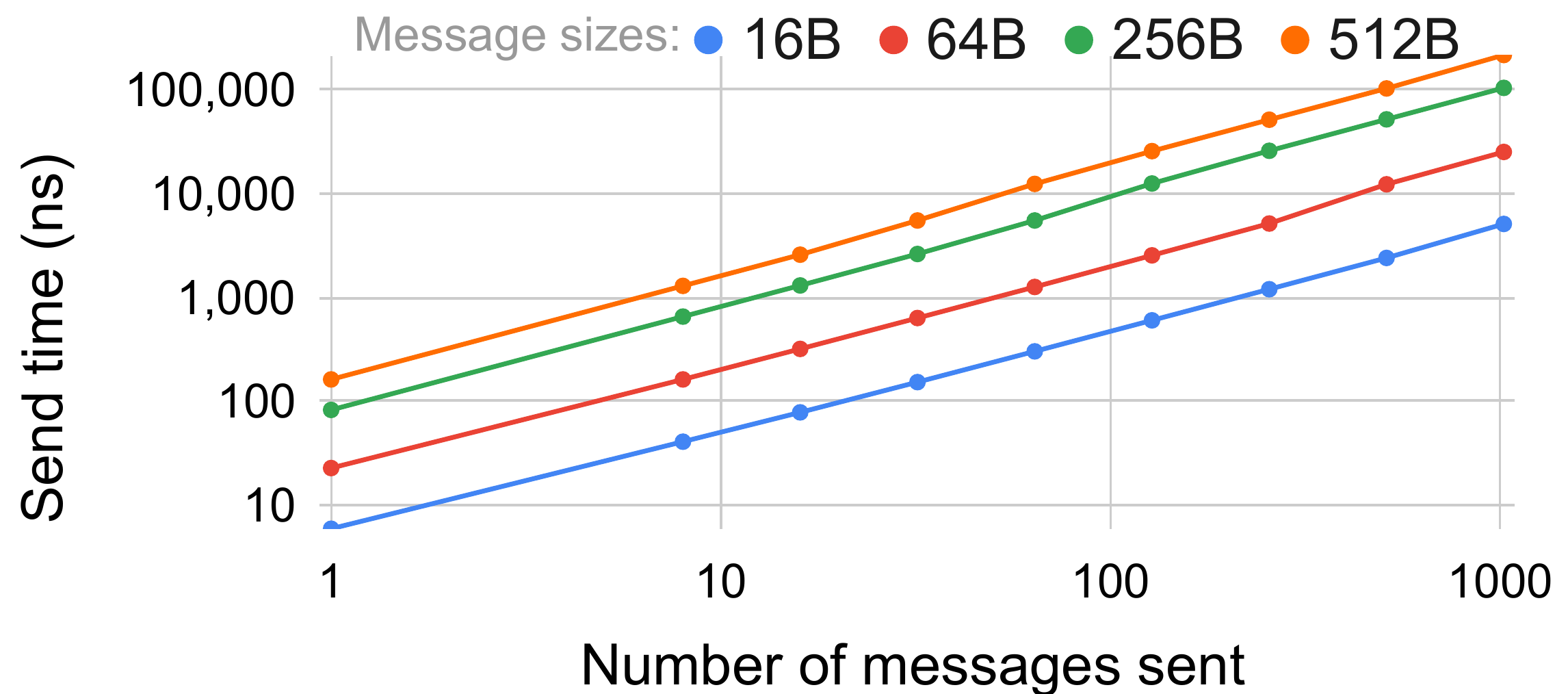}
    %\vspace{-20pt}
    \caption{Time to send increasing number of messages.}
    \label{fig:bench-send}
\end{figure}

\subsection{Sorting Benchmark}

We evaluate the effect of the parameters described above on the performance of distributed sorting algorithms. We start with a microbenchmark for sorting keys locally on a single core, then we evaluate an existing distributed sorting algorithm, MilliSort. Finally, we evaluate parameters in the design space of NanoSort, and present results from large scale simulations with 65,536 cores in the next section.

\subsubsection{Local Sort}
\begin{figure}[t]
\centering
    \includegraphics[width=0.6\linewidth]{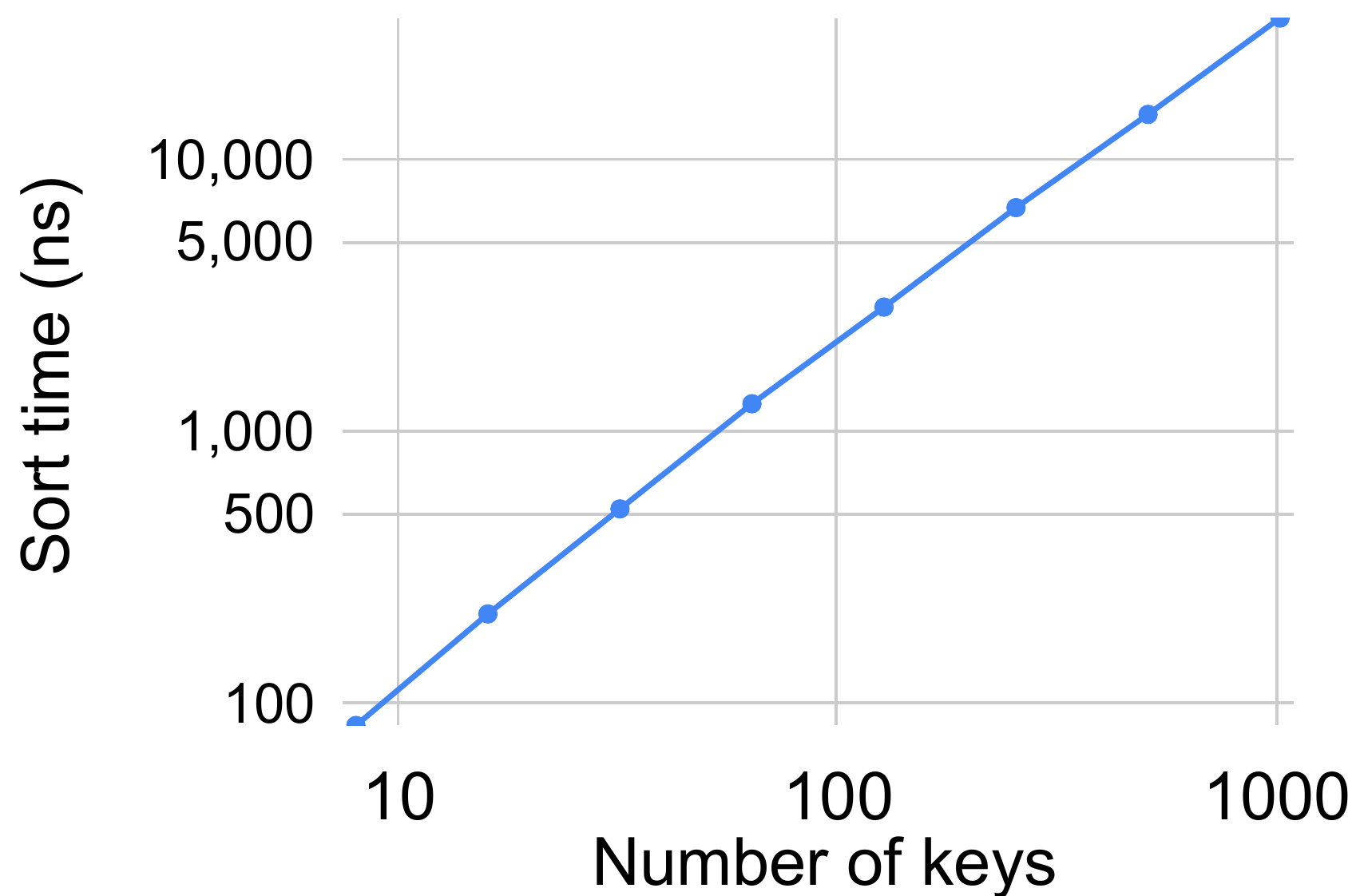}
    %\vspace{-20pt}
    \caption{Sorting locally with a single core.}
    \label{fig:bench-sort}
\end{figure}

Figure~\ref{fig:bench-sort} shows the time for a single RISC-V core to sort an increasing number of keys (integers). Each point is the average of 10 runs. We cleared the cache before each run, so that they would all experience the effects of the cache hierarchy. It takes over 30us to sort 1,024 keys, which is too long for a nanoTask. The appropriate number of keys for a nanoTask appears to be at most ~64 keys.

\subsubsection{MilliSort}
\begin{figure}[t]
\centering
    \includegraphics[width=0.5\linewidth]{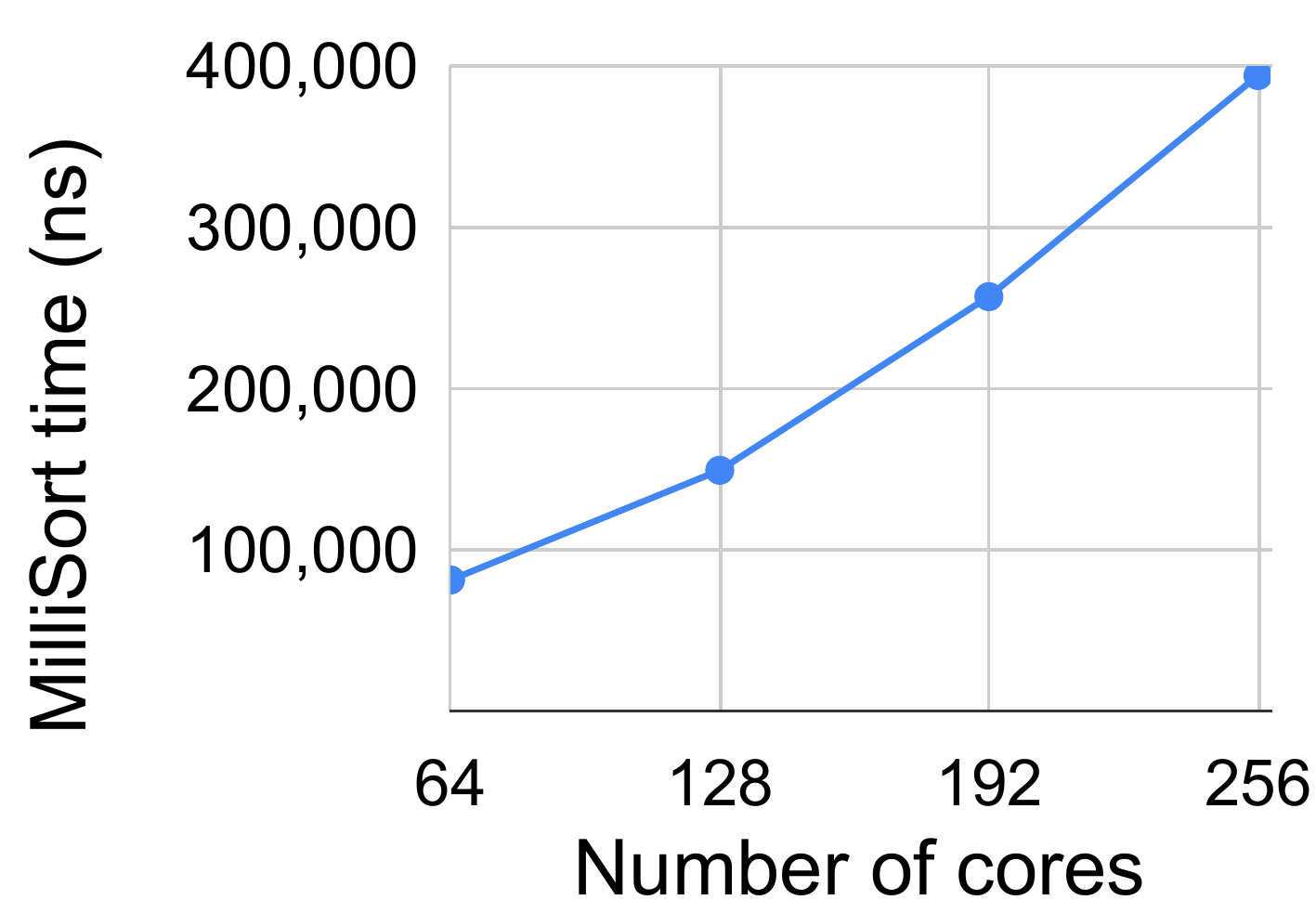}
    %\vspace{-20pt}
    \caption{MilliSort runtime on nanoPU with increasing number of cores (4,096 keys, 4 incast)}
    \label{fig:millisort-cores}
\end{figure}

Figure~\ref{fig:millisort-cores} shows the time for MilliSort to sort an increasing number of cores with 16 keys per core. The partitioning phase of the MilliSort algorithm picks the boundaries for each of the final buckets, which equals the number of cores; the more cores, the more bucket boundaries to pick. Thus, MilliSort's partition time increases with the number of cores, causing it to scale poorly to hundreds of cores. As we see in Figure~\ref{fig:millisort-cores}, MilliSort's runtime jumps from 61us with 64 cores, to 400us with 256 cores.

\begin{figure}[t]
\centering
    \includegraphics[width=0.5\linewidth]{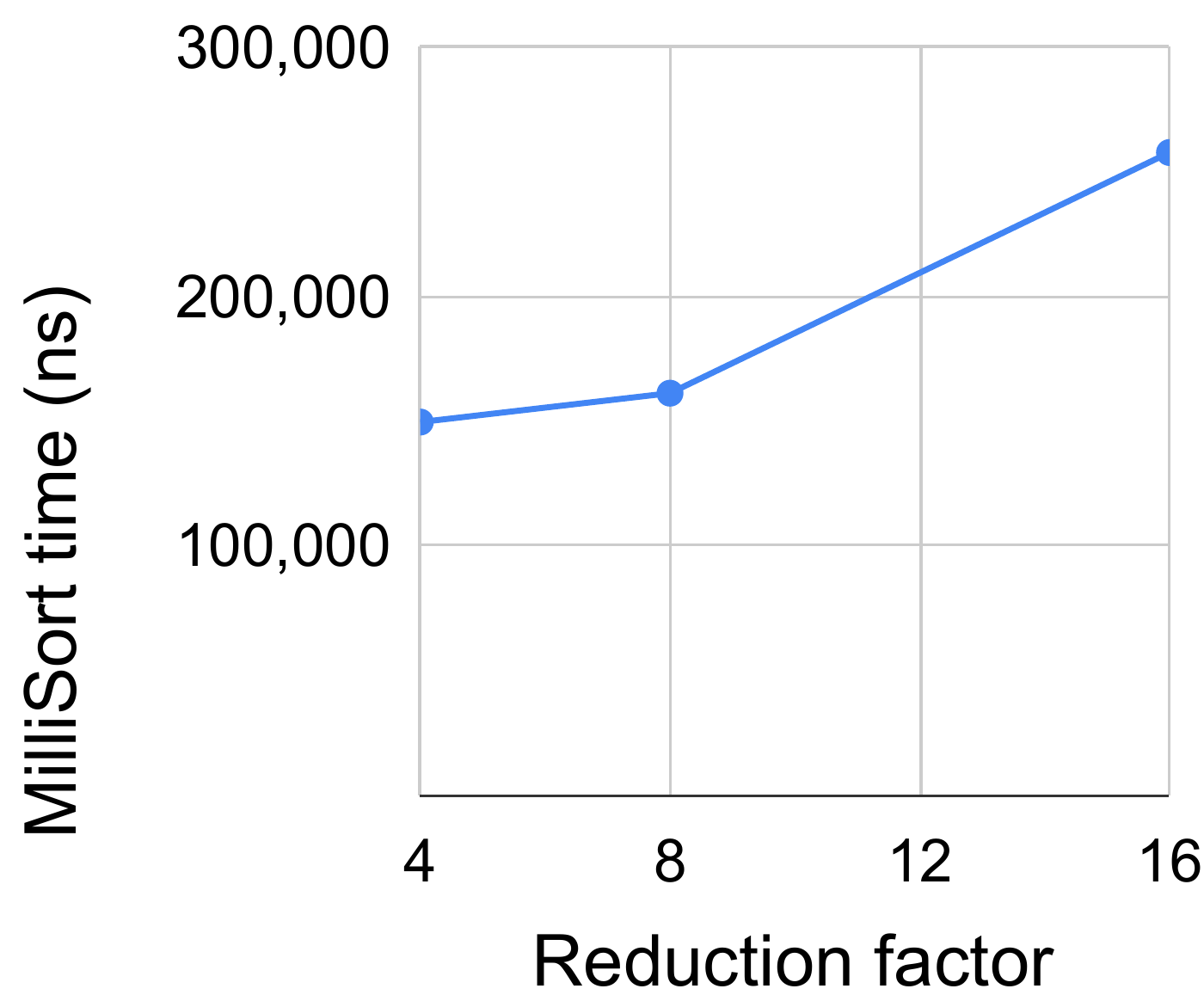}
    %\vspace{-20pt}
    \caption{MilliSort runtime on nanoPU with different incast sizes (128 cores, 4,096 keys)}
    \label{fig:millisort-incast}
\end{figure}

To decrease MilliSort's runtime, we tried varying its key parameter: the ``reduction factor'', which controls the number of pivot sorters per core (essentially, the incast). Figure~\ref{fig:millisort-incast} shows that increasing MilliSort's reduction factor causes a slowdown. This is because each pivot sorter must process a larger incast. 

\subsubsection{NanoSort}

We evaluate the effect of NanoSort ``knobs'', as well as the importance of network characteristics, including low switching latency, low tail latency and multicast support. Finally, we show the NanoSort runtime using 65,536 cores.

\steve{Should be consistent with terms parameters and knobs.}

\paragraph{Tuning NanoSort knobs.} 
Section~\ref{sec:design} describes ``knobs'' that affect the performance of NanoSort: the bucket size, keys per core and total number of cores.

\begin{figure}[t]
    \centering
    \begin{subfigure}[t]{.7\columnwidth}
        \centering
        \includegraphics[width=\columnwidth]{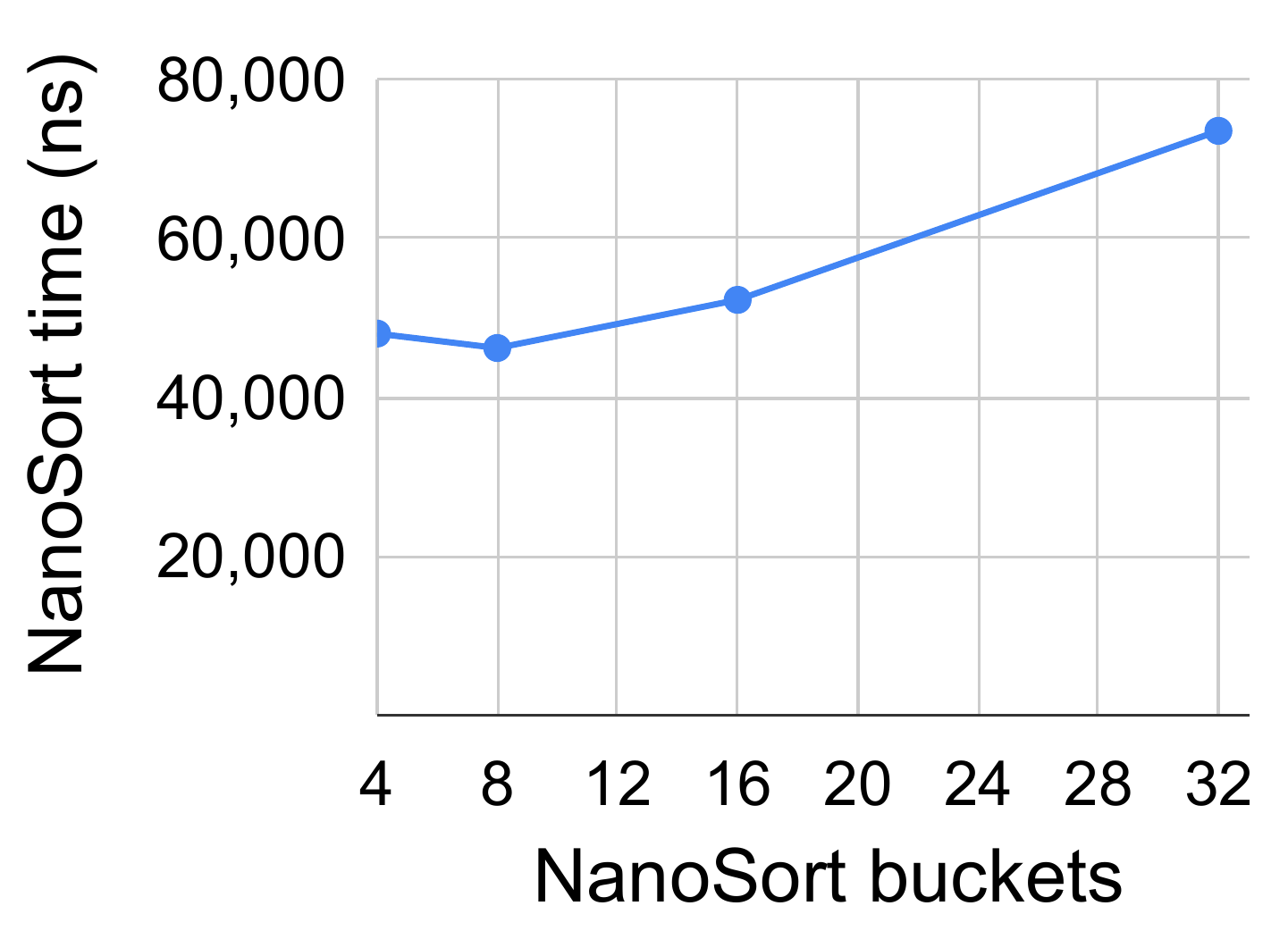}
        \vspace{-15pt}
        \caption{Runtime}
        \label{fig:nanosort-columns}
    \end{subfigure}
    \begin{subfigure}[t]{.7\columnwidth}
        \centering
        \includegraphics[width=\columnwidth]{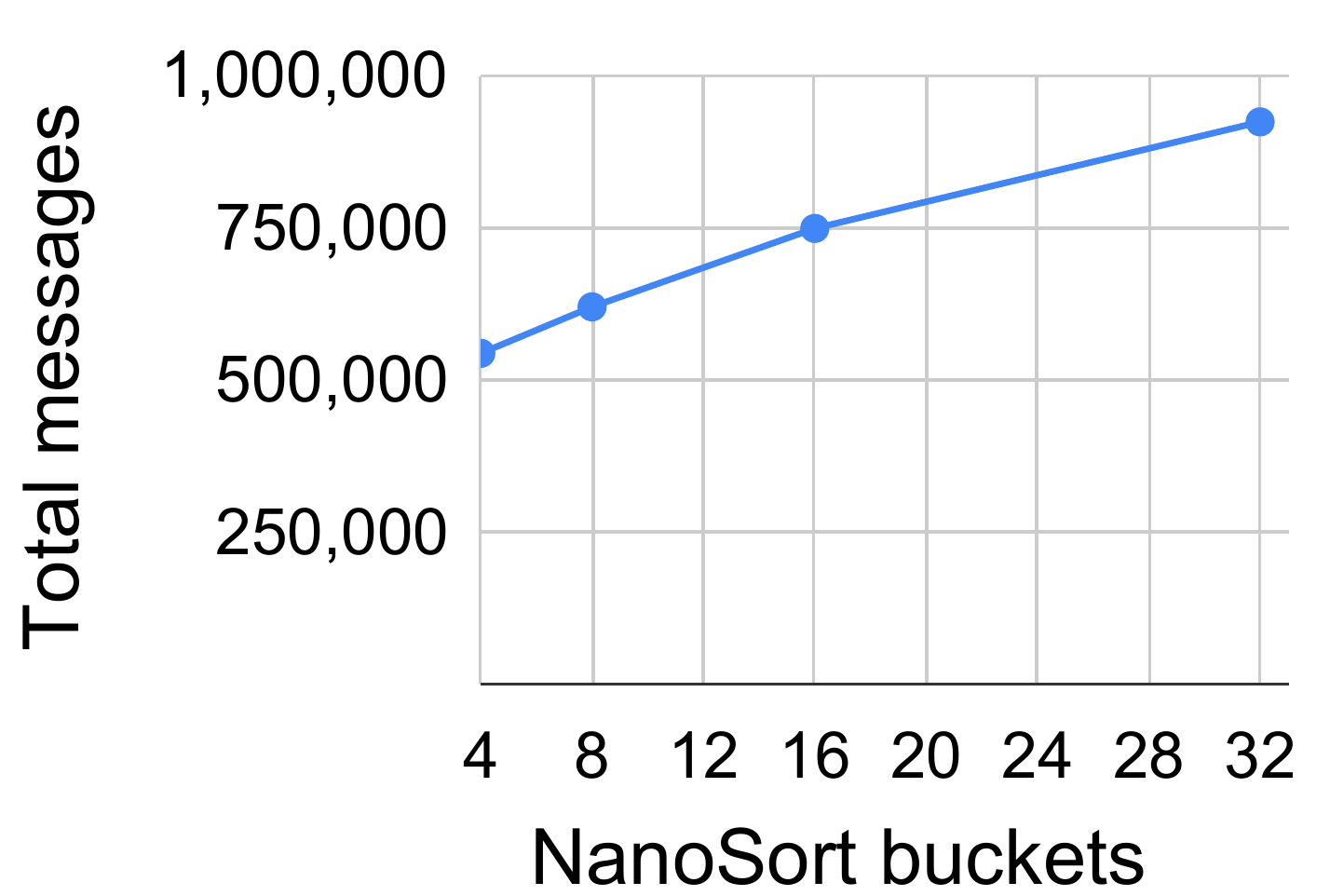}
        \vspace{-15pt}
        \caption{Traffic}
        \label{fig:nanosort-msgs}
    \end{subfigure}
    \caption{Effect of varying the number of buckets (4096 cores, 32 keys/core, 263ns switch latency).}
    \label{fig:nanosort-knobs}
\end{figure}

We ran NanoSort with a varying number of buckets, and fixed number of cores and keys per core. Figure~\ref{fig:nanosort-columns} shows that with 4,096 cores and 32 keys/core, using 4, 8 or 16 buckets has similar performance. This is despite the fact that they generate different amounts of network traffic (Figure~\ref{fig:nanosort-msgs}). This is because, as shown with the microbenchmarks in Section~\ref{sec:microbench}, there is a trade-off between the tree depth and width.

\begin{figure}[!ht]
    \centering
    \includegraphics[width=0.6\linewidth]{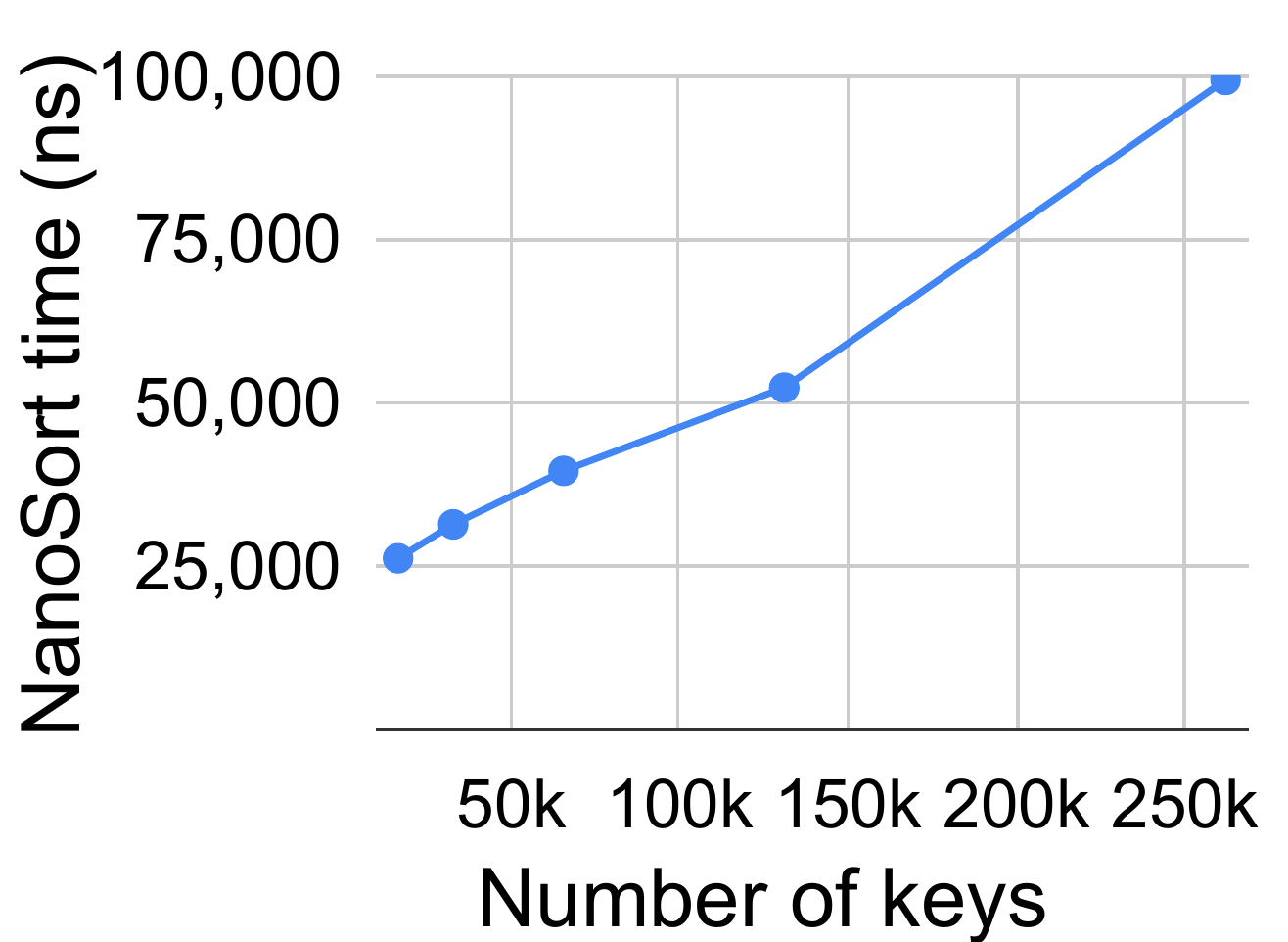}
    %\vspace{-20pt}
    \caption{Sorting an increasing number of keys with a fixed number of cores (4,096 cores, 263ns switch latency).}
    \label{fig:nanosort-keys}
\end{figure}

We ran NanoSort on 4,096 cores with an increasing number of keys, as shown in Figure~\ref{fig:nanosort-keys}. As the number of keys increases along the x-axis, as does the number of keys per core (e.g. 16K keys uses 4 keys/core, and 262K keys uses 64 keys/core). The runtime seems to increase linearly, which is because the work per core increases proportionally to the number of keys.

\begin{figure}[!ht]
    \centering
    \includegraphics[width=0.6\linewidth]{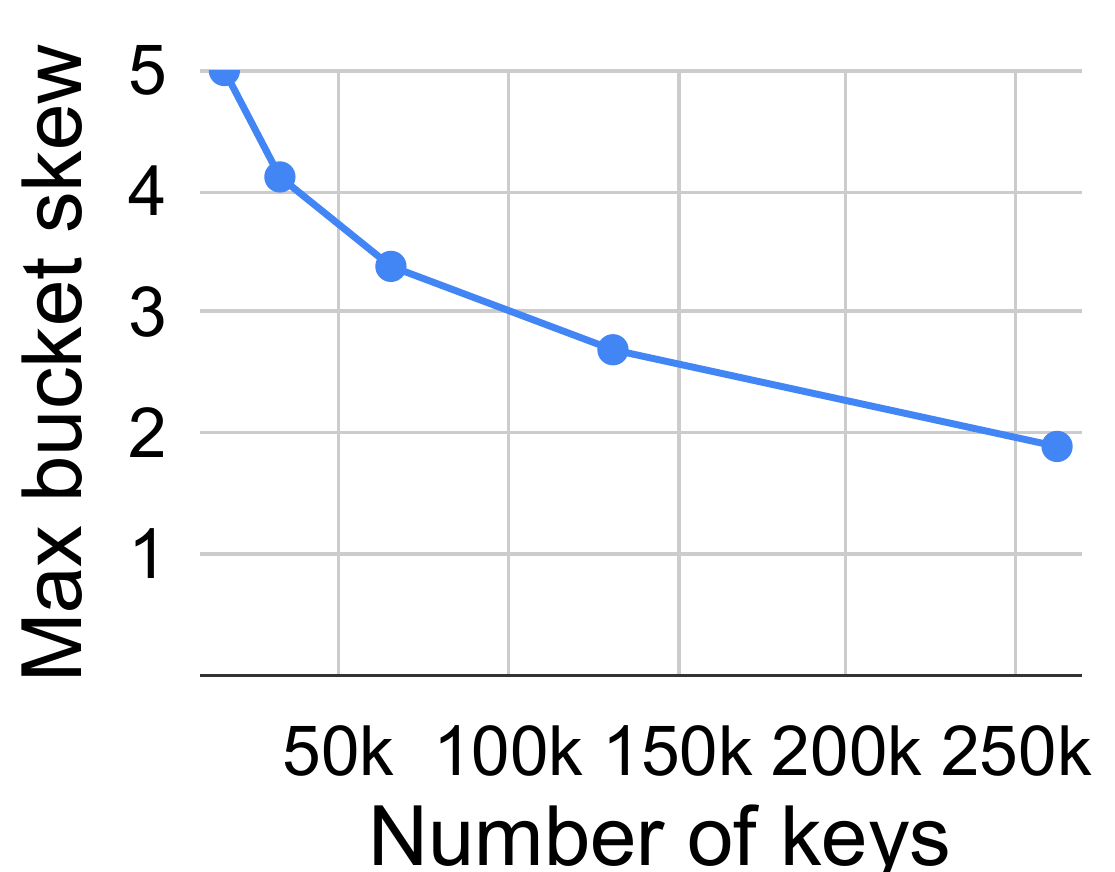}
    %\vspace{-20pt}
    \caption{Skew reduces with an increasing number of keys per core (4,096 cores, 263ns switch latency).}
    \label{fig:nanosort-skew}
\end{figure}

As the total number of keys increases, the maximum skew of the final buckets decreases (Figure~\ref{fig:nanosort-skew}). This is because the keys per core increases, so each core has better ``visibility'' of the distribution, and is able to pick better medians for the bucket boundaries.

\begin{table*}
\centering
\begin{tabular}{llrrrr}
\toprule
 & \textbf{CPU} & \textbf{Cores} & \textbf{SMT} & \textbf{1M sort} 
 \textbf{($\mu$s)} & \textbf{Tput (records/ms/core)}\\
\midrule
NanoSort     & RISC-V Rocket @ 3.2GHz  & 65,536 & 1 &   68 & 224 \\
MilliSort    & Xeon Gold 6148 @ 2.4GHz &  2,240 & 1 & 1000 & 1297 \\
TencentSort  & IBM POWER8 @ 2.9GHz     & 10,240 & 8 &   N/A & 1977 \\ % 512 nodes, 2x 10-core
CloudRAMSort & Xeon X5680 @ 2.9GHz     &  3,072 & 2 &   N/A & 707 \\ % 256 nodes, 2x 6-core. tput = ((1e12 / 100) / 4600 ) / (256 * 12)
%TritonSort   & Xeon E5-2670 v2 @ 2.5GHz&  5,952 & 2 & N/A & 1.22 \\ % 186 nodes, 32 vCPU. tput = ((1e12 / 100) / 1378000) / (186 * 32)
%TritonSort   & Intel E5520 @ 2.27 GHz  & 416  & 2 &   N/A &  \\ % 52 nodes, 2x 4-core. tput = (( / 100) /  ) / (52 * 8)
\bottomrule                                                              
\end{tabular}
\caption{Per-core efficiency comparison. The throughputs are measured from sorting different numbers of records on each system: 1M with NanoSort, 26M with MilliSort, 10B with TencentSort and CloudRAMSort.}
\label{tab:throughput}
\end{table*}

\paragraph{Switching latency.}

\begin{figure}[t]
    \centering
    \begin{subfigure}[t]{.7\columnwidth}
        \centering
        \includegraphics[width=\columnwidth]{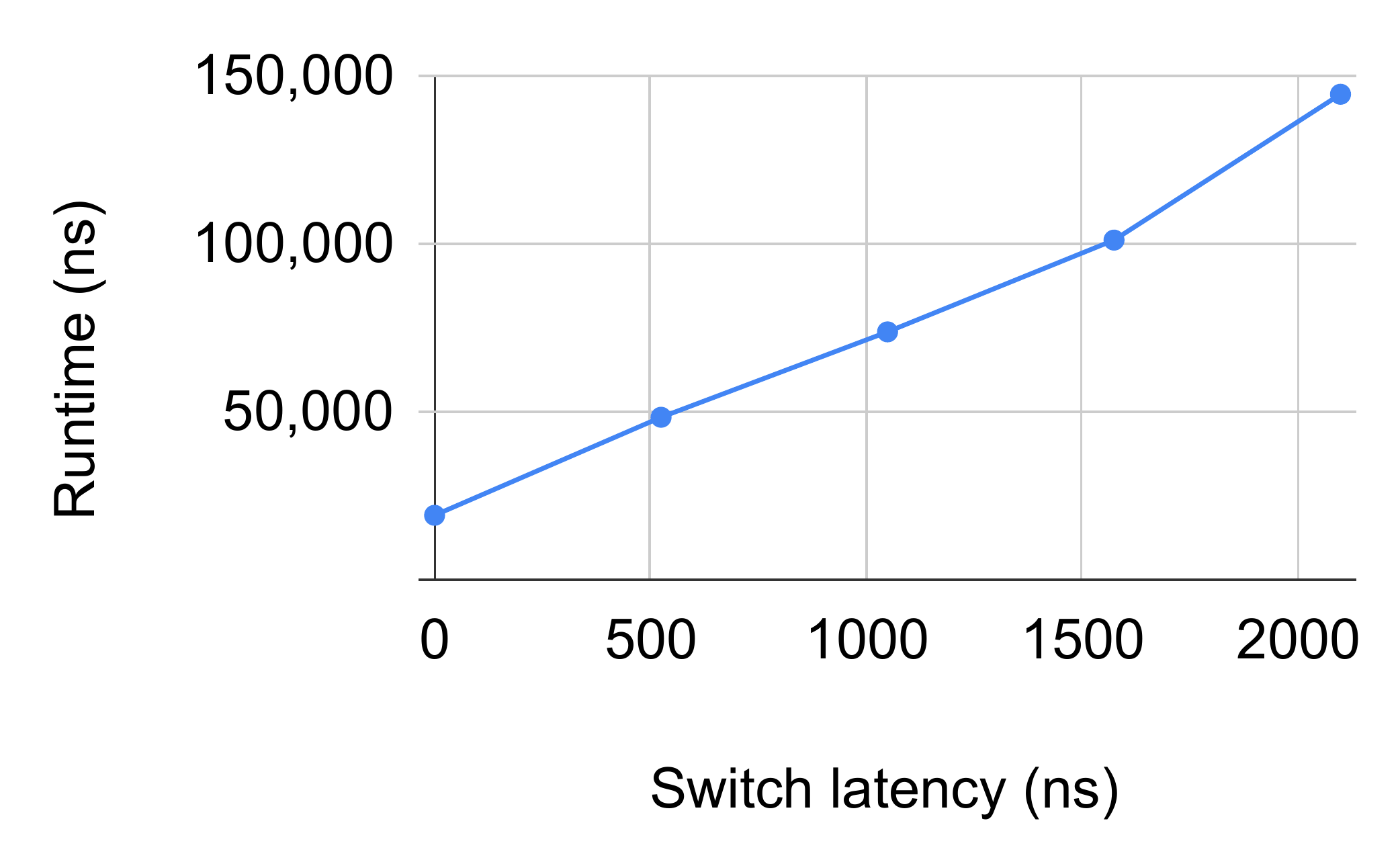}
        \vspace{-15pt}
        \caption{Runtime}
    \end{subfigure}
    \begin{subfigure}[t]{.7\columnwidth}
        \centering
        \includegraphics[width=\columnwidth]{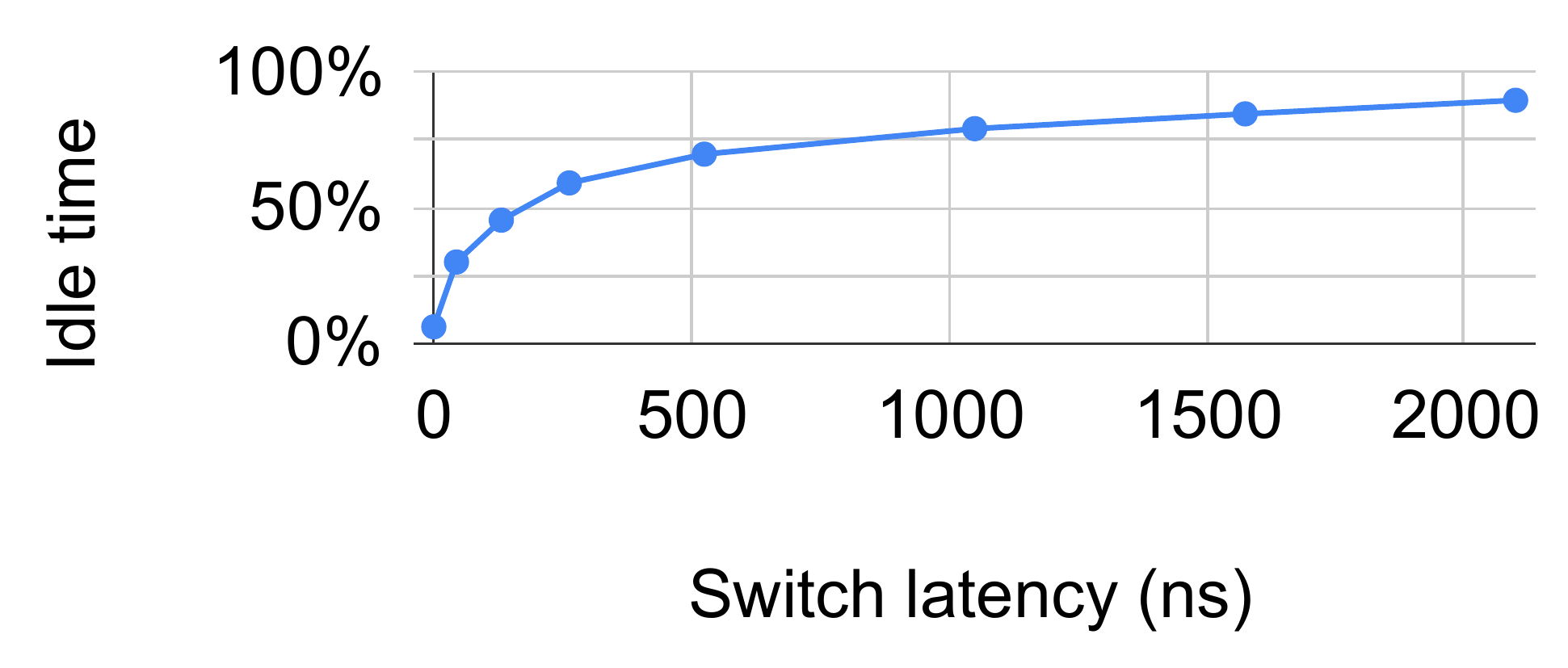}
        \vspace{-15pt}
        \caption{Idle time}
        \label{fig:nanosort-idle}
    \end{subfigure}
    \caption{Effect of switching latency on NanoSort (64 cores, 16 keys/core).
    %\theo{should we run this with 4,096 cores? I think it's a stronger result that high latency hurts the performance with only a few cores, let alone thousands of cores.}
        }
    \label{fig:nanosort-switchlat}
\end{figure}

% cols=16, keysPerNode=16
Figure~\ref{fig:nanosort-switchlat} shows the effect of the switching latency on NanoSorting 1K keys with 64 cores. As expected, as the switching latency increases, so does the runtime. This is because the cores spend more time idling, waiting for messages, as we can see in Figure~\ref{fig:nanosort-idle}.

\begin{figure}[t]
    \centering
    \includegraphics[width=0.6\linewidth]{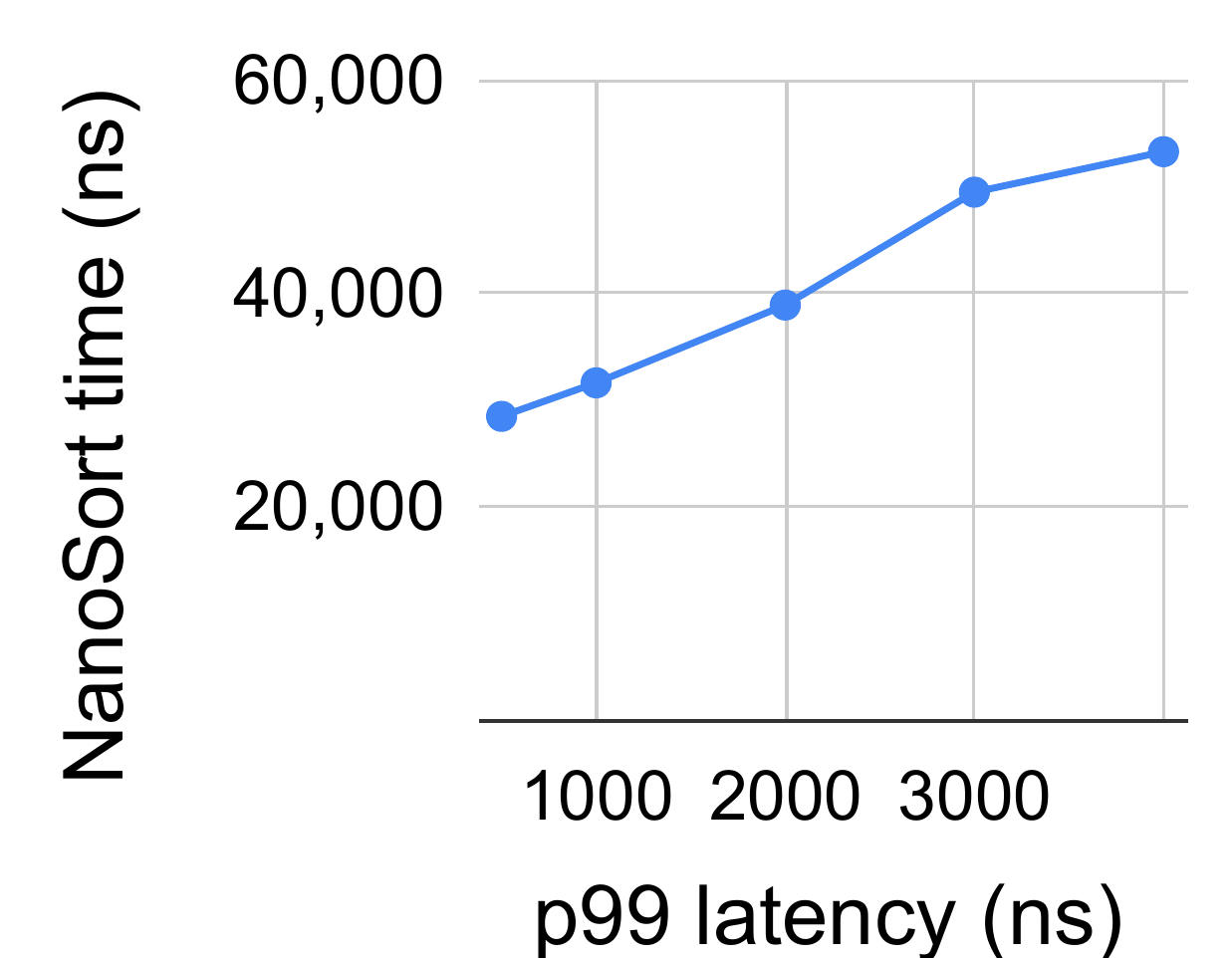}
    %\vspace{-20pt}
    \caption{Effect of tail latency (256 cores, 8 buckets, 32 keys/core, 263ns switch latency)}
    \label{fig:nanosort-tail}
\end{figure}

\paragraph{Tail latency.} Not only is the minimum communication time (switching latency) important, but also the tail. We ran NanoSort on 256 cores with 131K keys, and injected an additional latency to 1\% of messages (i.e.\ the 99th percentile tail latency). Figure~\ref{fig:nanosort-tail} shows how increasing this 99th percentile latency affects runtime. With a p99 latency of 4000ns, NanoSort takes 53$\mu$s, double the baseline time of 26$\mu$s.
When so many messages are sent (38K messages in this experiment) it is inevitable that the tail latency will be experienced.
This demonstrates that it is crucial that the latency of the tail be minimized.

% For mcast experiment: columns=16
\paragraph{Multicast support.} We ran NanoSort on 4,096 cores with and without multicast support in the network. Without multicast, when a core picks a median, it must individually send it to all other nodes in its group. In the first level of recursion, this means sending the same message to all other 4,096 nodes. The the effect is clear: without multicast, NanoSort takes 96$\mu$s; adding multicast support reduces the number of messages sent by 18\%, and the runtime to 40$\mu$s (2.4x speedup).

\subsection{Datacenter-Scale NanoSorting}
To see whether NanoSort can scale to thousands of cores, we ran a large scale simulation of 65,536 nanoPU cores using 4,224 AWS EC2 vCPUs. We measured the time for NanoSort to sort 1M keys (16 keys per node and 16 buckets). Of 10 runs, all took less than 78$\mu$s, with an average time of 68$\mu$s (4.127$\mu$s standard deviation). For comparison, this is an order of magnitude faster than MilliSort running on a HPC cluster, which sorts 0.9M keys in 1ms~\cite{millisort}. 
The state of the art sorting algorithm for multicore systems, IPS$^4$o, sorts 1M keys in 2ms~\cite{ips40}.

%\begin{figure}[!ht]
%    \centering
%    \includegraphics[width=0.6\linewidth]{idle-time}
%    %\vspace{-20pt}
%    \caption{Idle time for cores in large-scale NanoSort}
%    \label{fig:idle-time}
%\end{figure}
%\begin{figure}[!ht]
%    \centering
%    \includegraphics[width=0.6\linewidth]{nanosort-breakdown}
%    %\vspace{-20pt}
%    \caption{Breakdown of time spent by each core on the stages of NanoSort}
%    \label{fig:breakdown}
%\end{figure}
\begin{figure}[t]
    \centering
    \begin{subfigure}[t]{.6\columnwidth}
        \centering
        \includegraphics[width=\columnwidth]{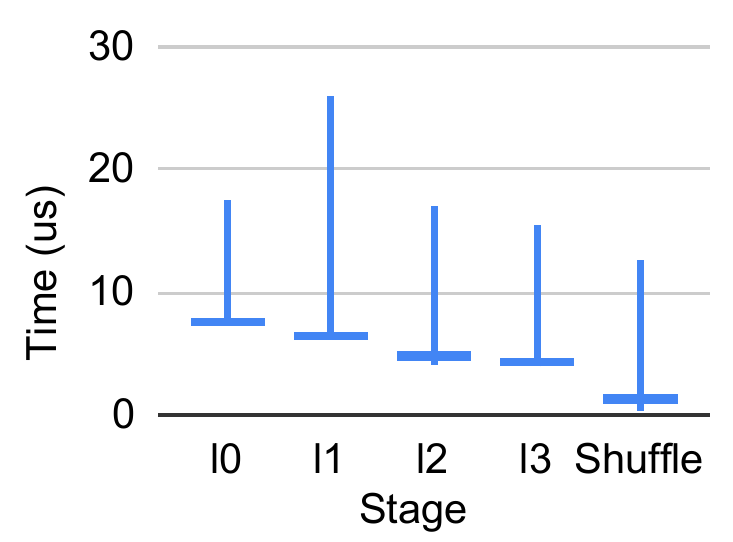}
        \vspace{-15pt}
        \caption{Execution time}
        \label{fig:breakdown}
    \end{subfigure}
    \begin{subfigure}[t]{.6\columnwidth}
        \centering
        \includegraphics[width=\columnwidth]{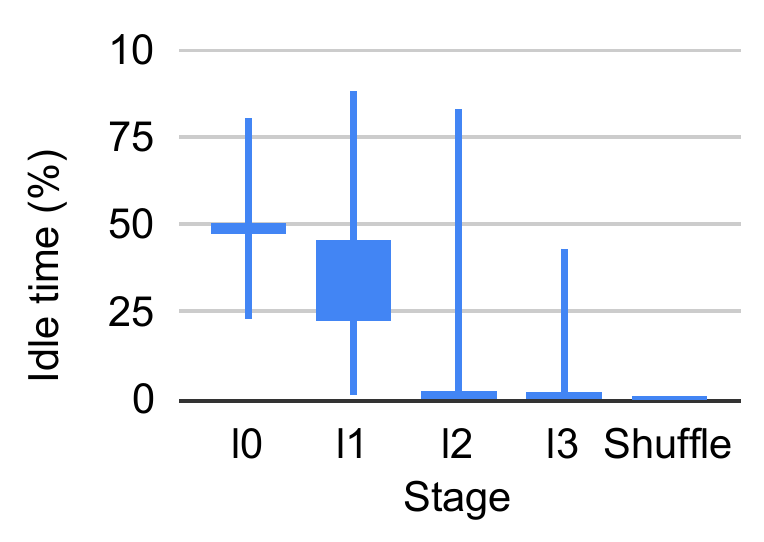}
        \vspace{-15pt}
        \caption{Idle time}
        \label{fig:idle-time}
    \end{subfigure}
    \caption{Execution breakdown for NanoSort with 65,536 cores.}
\end{figure}

\paragraph{Execution breakdown}
Figure~\ref{fig:breakdown} shows the distribution of time spent on each stage for all the cores. We can see that there is some variance, but it is modest.
The first stage (level of recursion 0) is the fastest. It also has the least variance, because all the workers start at the same time, have exactly the same number of initial keys, and perform the same operation (send/receive each key).
The variance in the other stages is not due to compute time, but the time the core spends idle, waiting for messages from other cores (possibly still in previous stages).
Figure~\ref{fig:idle-time} shows the distribution of idle times for the cores on each stage.
There is less variance in idle time in the first stages. In the last stage, shuffling, the cores are sending and receiving the final values, so they are constantly active until completion.

\paragraph{Efficiency}
Table~\ref{tab:throughput} compares the sorting throughput of NanoSort to MilliSort and other systems in the GraySort benchmark.
There are multiple reasons why NanoSort has a lower per-core throughput than the other systems: it is sorting fewer records (1M) than the other systems; it has wimpier cores without SMT (8x less than TencentSort); and it is using many more cores. There is a trade-off between latency and throughput: sorting with a tight time budget comes at the expense of reduced efficiency.

% old figs:
% \begin{figure}
%     \centering
%     \includegraphics[width=0.8\linewidth]{nanosort-keys-per-node}
%     %\vspace{-20pt}
%     \caption{Effect of keys/core on execution time (256 cores, 263ns switch latency). The number of columns equals the keys/core for each round. }
%     \label{fig:nanosort-keys-per-node}
% \end{figure}

%-------------------------------------------------------------------------------
\section{Related Work}
%-------------------------------------------------------------------------------
\label{sec:related}

NanoSort explores a new extreme of granular computing using new low-latency communication designs and sorting algorithms.

\paragraph{Distributed sorting algorithms.}
Parallel sorting algorithms were designed for multi-core systems, and were initially limited by disk I/O~\cite{aggarwal1988io,vitter1985bucket,vitter2008algorithms}.
AlphaSort~\cite{nyberg1995alphasort} addresses this bottleneck with cache-sensitive external sort.
Various sampling-based techniques have been proposed for sorting on a single machine:
IPS$^4$o~\cite{ips40} describes a distribution sort optimized for multi-core machines and LearnedSort~\cite{learnedsort} creates a hierarchecal model to determine bucket boundaries.
Manycore GPUs can be used to accelerate radix and merge sorts~\cite{satish2009gpu}.
Some systems are designed for high-throughput (as opposed to low-latency):
CloudRAMSort~\cite{cloudramsort}, TritonSort~\cite{tritonsort} and TencentSort~\cite{tencentsort} perform large scale sort using hundreds of machines.

%\begin{itemize}
%    \item Sorting used to be limited by disk I/O: The Input/Output Complexity of Sorting and Related Problems~\cite{aggarwal1988io}. "sorting continues to account for roughly one-fourth of all computer cycles."
%    \item In-Place Parallel Super Scalar Samplesort (IPS4o)~\cite{ips40}
%    \item Theoretically-Efficient and Practical Parallel In-Place Radix Sorting
%    \item Designing Efficient Sorting Algorithms for Manycore GPUs~\cite{satish2009gpu}
%    \item CloudRAMSort: fast and efficient large-scale distributed RAM sort on shared-nothing cluster
%    \item TritonSort: A Balanced Large-Scale Sorting System (NSDI’11)~\cite{tritonsort}
%    \item Tencent Sort
%    \item The Case for a Learned Sorting Algorithm~\cite{learnedsort}
%    \item Parallel Sorting by Regular Sampling
%    \item Algorithms and Data Structures for External Memory
%\end{itemize}

\paragraph{Low-latency communication.}
eRPC~\cite{eRPC} reduces RTT with off-the-shelf NICs.
The nanoPU~\cite{nanopu} and NeBULA~\cite{nebula} prototype new low-latency NIC/CPU codesigns.
While NeBULA delivers messages to the cache, the nanoPU delivers messages straight to the CPU register file, with has a lower latency.
Shinjuku~\cite{shinjuku}, Shenango~\cite{shenango} and Zygos~\cite{zygos} explore OS scheduling to reduce request latency.

\paragraph{Granular computing.}
Our work addresses the questions raised by Lee et.\ al in their position paper on granular computing~\cite{lee19}.
Other work has investigated increasing parallelism~\cite{ExCamera,gg,PyWren} for interactive applications and for handling ``bursts'', like millisort~\cite{millisort}.
%Collin Lee's dissertation on Distributed Procedure Calls~\cite{lee2021dpc}:
%\url{
%https://web.stanford.edu/~ouster/cgi-bin/papers/LeePhD.pdf
%}

%-------------------------------------------------------------------------------
\section{Conclusion}
\label{sec:conclusion}
%-------------------------------------------------------------------------------

The datasets processed by real-time systems is growing in size. If they are to continue providing interactive results, they must scale out, increasing parallelism and in turn, granularity.
As granularity increases, the communication overhead and coordination dominates the processing time.
We show how new network design and new algorithms can be leveraged to reduce the communication overhead and scale out with the dataset.

%-------------------------------------------------------------------------------
\bibliographystyle{plain}
\bibliography{references}

\appendix

%\input{text/nanopu}

%%%%%%%%%%%%%%%%%%%%%%%%%%%%%%%%%%%%%%%%%%%%%%%%%%%%%%%%%%%%%%%%%%%%%%%%%%%%%%%%
\end{document}